\RequirePackage{rotating}
\documentclass[fleqn,usenatbib]{mnras}
\usepackage{graphicx,bm,times}
\usepackage{subfig}
\usepackage{float}
\usepackage{import}
\usepackage{multirow}
\usepackage{natbib}
\usepackage[labelfont=bf]{caption}
\usepackage{newtxtext,newtxmath}
\usepackage{rotating}
\usepackage[T1]{fontenc}
\usepackage{hyperref}
\usepackage{natbib}
\usepackage{amsmath}
\usepackage{graphicx}
\usepackage{multirow}
\usepackage{url}
\usepackage{enumitem}
\usepackage[table]{xcolor}
\usepackage{bold-extra}

\graphicspath{{Images/}} 

\setcitestyle{authoryear,open={(},close={)}}
\DeclareRobustCommand{\VAN}[3]{#2}
\let\VANthebibliography\thebibliography
\def\thebibliography{\DeclareRobustCommand{\VAN}[3]{##3}\VANthebibliography}

\begin{document}

\author[O.Thompson et al.]{Oliver A. Thompson$^{1,2}$\thanks{Contact e-mail: \href{O.Thompson@hull.ac.uk}{O.Thompson@hull.ac.uk}}, 
Alexander J. Richings$^{1,2}$,
Brad K. Gibson$^{1}$, 
Claude-André Faucher-Giguère$^{3}$,
\newauthor
Robert Feldmann$^{4}$,
Christopher C. Hayward$^{5}$
\\
$^1$E.A. Milne Centre for Astrophysics, University of Hull, Hull, HU6 7RX, UK 
\\
$^2$Centre of Excellence for DAIM, University of Hull, Hull, HU6 7RX, UK
\\
$^3$Department of Physics \& Astronomy and CIERA, Northwestern University, 1800 Sherman Ave, Evanston, IL 60201, USA
\\
$^4$Institute for Computational Science, University of Zurich, Winterthurerstrasse 190, Zurich CH-8057, Switzerland
\\
$^5$Center for Computational Astrophysics, Flatiron Institute, 162 Fifth Avenue, New York, NY 10010, USA
}

\title[Predictions for CO emission in simulations]{Predictions for CO emission and the CO-to-H$_2$ conversion factor in galaxy simulations with non-equilibrium chemistry}
\date{\today}
\pagerange{\pageref{firstpage}--\pageref{lastpage}}
\maketitle
\pagenumbering{arabic}
\newpage
\label{firstpage}

\begin{abstract}
Our ability to trace the star-forming molecular gas is important to our understanding of the Universe. We can trace this gas using CO emission, converting the observed CO intensity into the H$_2$ gas mass of the region using the CO-to-H$_2$ conversion factor ($X_{\mbox{\textsc{co}}}$). In this paper, we use simulations to study the conversion factor and the molecular gas within galaxies. We analysed a suite of simulations of isolated disc galaxies, ranging from dwarfs to Milky Way-mass galaxies, that were run using the FIRE-2 subgrid models coupled to the CHIMES non-equilibrium chemistry solver. We use the non-equilibrium abundances from the simulations, and we also compare to results using abundances assuming equilibrium, which we calculate from the simulation in post-processing. Our non-equilibrium simulations are able to reproduce the relation between CO and H$_2$ column densities, and the relation between $X_{\mbox{\textsc{co}}}$ and metallicity, seen within observations of the Milky Way. We also compare to the xCOLD GASS survey, and find agreement with their data to our predicted CO luminosities at fixed star formation rate. We also find the multivariate function used by xCOLD GASS overpredicts the H$_2$ mass for our simulations, motivating us to suggest an alternative multivariate function of our fitting, though we caution that this fitting is uncertain due to the limited range of galaxy conditions covered by our simulations. We also find that the non-equilibrium chemistry has little effect on the conversion factor (<5\%) for our high-mass galaxies, though still affects the H$_2$ mass and $L_{\mbox{\textsc{co}}}$ by $\approx$25\%.

\end{abstract}


\section{Introduction} \label{Introduction}
Molecular gas plays a vital role within the formation of galaxies, as it fuels the formation of new stars. However, one of the most abundant molecules, H$_2$, is difficult to detect directly as it does not produce emission at the cold temperatures (10-20K) typical of the dense molecular clouds that contain most of a galaxy's reservoir of molecular gas. For that reason, we have to rely on using other methods to detect and measure the molecular gas content within galaxies. Some methods use IR-based emission, and assume some dust-to-gas ratio to calculate the mass of H$_2$, whilst others use emission from the molecular carbon monoxide, or CO.

CO is used to trace the molecular gas as it has its first transition (J=1-0) at \textasciitilde5.5 K, which can be easily excited at the typical temperatures found within molecular clouds. CO molecules within these molecular clouds can be excited through a combination of collisions with H$_2$ and radiative trapping, and de-excite spontaneously through collisions and emissions. However, CO is not a perfect tracer for H$_2$, as it is more easily dissociated, resulting in a `dark gas' \citep{van_dishoeck_photodissociation_1988, wolfire_dark_2010}. This CO-dark molecular gas emits CO weakly and it has been estimated that more than 30\% of the molecular H$_2$ gas is CO-dark using gamma ray emission within the Milky Way \citep{grenier_unveiling_2005}, with evidence of CO-dark gas in various extragalactic studies too \citep{poglitsch_multiwavelength_1995,madden_c_1997}. Suggestions for tracers of this faint-CO  within the Milky Way have been proposed, such as OH and HCO+ \citep{liszt_galactic_1996,allen_structure_2015,nguyen_dustgas_2018}. More general suggestions that can be used outside the Milky Way have included atomic carbon \citep{gerin_atomic_2000,papadopoulos_c_2004,offner_alternative_2014,glover_modelling_2015,li_dark_2018,clark_tracing_2019}, [C \textsc{ii}] \citep{zanella_c_2018,madden_tracing_2020,vizgan_tracing_2022,deugenio_cii_2023,ramambason_modeling_2024}, and in a small range of environments, HCl \citep{schilke_hydrogen_1995}. Due to a significant amount of H$_2$ being undetectable by CO, we require a way to infer the H$_2$ mass from the CO we can observe.

We can link observed line emissions from CO to the total H$_2$ mass using the CO-to-H$_2$ conversion factor, $X_{\mbox{\textsc{co}}}$, which is defined as:

\begin{equation} \label{XCO_eqn}
X_{\hbox{\textsc{co}}} = \frac{N_{\mbox{\textsc{h}}_{\oldstylenums{2}}}}{W_{\hbox{\textsc{co}}}}, 
\end{equation}

where $N_{\mbox{\textsc{h}}_{\oldstylenums{2}}}$ is the column density of H$_2$, and $W_{\mbox{\textsc{co}}}$ is the velocity-integrated CO(1-0) brightness temperature, which is related to the intensity of CO(1-0) emission line. We can also calculate an alternate version of the CO-to-H$_2$ conversion factor known as $\alpha_{\mbox{\textsc{co}}}$, which is defined as:

\begin{equation} \label{alpCO_eqn}
\alpha_{\mbox{\textsc{co}}}=\frac{M_{\mbox{\textsc{h}}_{\oldstylenums{2}}}}{L_{\mbox{\textsc{co}}}},
\end{equation}

where $M_{\mbox{\textsc{h}}_{\oldstylenums{2}}}$ is the H$_2$ gas mass, $L_{\mbox{\textsc{co}}}$ is the CO luminosity from the galaxy, meaning $\alpha_{\mbox{\textsc{co}}}$ is simply a mass-to-light ratio. 

\citet{bolatto_co--h_2013} and references therein showed that in the Milky Way, the $X_{\mbox{\textsc{co}}}$ conversion factor is estimated to be 2 $\times$ 10$^{20}$cm$^{-2}$(K km s$^{-1})^{-1}\pm$30\%, and the corresponding $\alpha_{\mbox{\textsc{co}}}$ is 4.3M$_\odot$ (K km s$^{-1}$ pc$^2$)$^{-1}\pm$30\% . These dust-based approaches are able to estimate the H$_2$ mass through IR emission from dust, allowing them to calculate the H$_2$ column density, assuming some dust-to-gas (DTG) ratio. These observations also showed a correlation between metallicity and $X_{\mbox{\textsc{co}}}$ \citep[see][Figure 9]{bolatto_co--h_2013} where as metallicity increases, the observed $X_{\mbox{\textsc{co}}}$ decreased, with a sharp increase with metallicities below $Z$ $\sim$ 1/3 - 1/2 $Z_\odot$. Observational studies are able to use $X_{\mbox{\textsc{co}}}$ and $\alpha_{\mbox{\textsc{co}}}$ to calculate the molecular gas mass from CO observations. Without a good knowledge and understanding of the conversion factor and how it depends on environment, observational studies would be unable to reasonably estimate the molecular gas content of local and distant galaxies.



It can be shown that $X_{\mbox{\textsc{co}}}$ varies within different galaxy populations by looking at the Kennicutt-Schmidt (KS) relation. When applying a constant $X_{\mbox{\textsc{co}}}$ to the KS law, metal-poor galaxies ($Z < 0.3Z_\odot$) do not line up with the main relation (see also \citeauthor{cormier_molecular_2014} \citeyear{cormier_molecular_2014}) and require higher values of $X_{\mbox{\textsc{co}}}$ to line up \citep[see][Figure 11]{kennicutt_star_2012}. For starburst galaxies, we also see a bimodal KS relation when applying a bimodal $X_{\mbox{\textsc{co}}}$. To achieve a unimodal relation, we require a continuous $X_{\mbox{\textsc{co}}}$ \citep[see][Figure 7]{narayanan_general_2012}. We also see the non-universality in $X_{\mbox{\textsc{co}}}$ in ULIRGs, where a Milky Way-like conversion factor would infer a gas mass that would exceed the dynamical mass of the galaxy \citep{solomon_molecular_1997}. 

Studies have attempted to create a function for the conversion factor dependant on metallicity \citep{wilson_metallicity_1995,schruba_low_2012,genzel_metallicity_2012} due to the relation mentioned prior. However, these prescriptions have a large scatter on the predicted conversion factor, and there are inconsistencies across these studies. For this reason, we require a multivariate function that captures additional dependencies that metallicity alone can not.

The xCOLD GASS survey \citep{saintonge_xcold_2017} observed CO emission from 532 galaxies and then used those observations to calculate various properties of those galaxies, including the molecular gas mass, $M_{\mbox{\textsc{h}}_{\oldstylenums{2}}}$. Rather than use the value stated in \citet{bolatto_co--h_2013}, this molecular gas mass was calculated by using the multivariate form of $\alpha_{\mbox{\textsc{co}}}$ from \citet{accurso_deriving_2017}, which depends on the local conditions of the galaxy as follows:

\begin{equation} \label{Accurso_Eqn}
\begin{split}
    \log \alpha_{\mbox{\textsc{co}}}(Z) = 0.742\log\frac{L_{\mbox{\textsc{[C} {\textsc{ii}}\textsc{]}}}}{L_{\mbox{\textsc{co}}(1-0)}} - 0.944[12 + \log(\mbox{O/H})] \\
- 0.109\log\Delta(\mbox{MS}) + 6.439,
\end{split}
\end{equation}

where $L_{\mbox{[C \textsc{ii}]}}$ is the observed [C \textsc{ii}] 158$\mu$m emission line luminosity, $\Delta(\mbox{MS})$ is the offset from the main sequence, and 12+log(O/H) is the metallicity. This model was found by fitting observational data from the \textit{Hershel} PACS observations of the [C \textsc{ii}] emission line \citep{poglitsch_photodetector_2010,pilbratt_herschel_2010}, and low-metallicity galaxies from the xCOLD GASS survey. Using observed scaling relations between the $L_{\mbox{\textsc{[C} {\textsc{ii}]}}}$/$L_{\mbox{\textsc{co}}(1-0)}$ ratio as a function of integrated properties, Bayesian analysis revealed metallicity and MS offset as the only two parameters needed to quantify the variations in the luminosity ratio. Combining these variations with a linear fit to their radiative transfer models gave the formula for a multivariate $\alpha_{\mbox{\textsc{co}}}$. We discuss the different forms of the \citet{accurso_deriving_2017} multivariate conversion factor in more detail in Section \ref{MultivariateSubsec}.

Other prescriptions for $X_{\mbox{\textsc{co}}}$ have been derived, relying on other galaxy parameters. \citet{teng_star_2024} uses dust-based measurements to calculate $\alpha_{\mbox{\textsc{co}}}$ for objects in the PHANGS-ALMA survey \citet{leroy_phangsalma_2021}, and then derived a prescription for $\alpha_{\mbox{\textsc{co}}}$ that uses velocity-dispersion. Similarly, \citet{hirashita_effects_2023} derive a prescription for $X_{\mbox{\textsc{co}}}$ that uses the dust-to-gas ratio and the surface area-to-mass ratio of the dust, which reflects the effect of the dust grain distribution. \citet{shetty_modelling_2011,shetty_modelling_2011-1} investigated $X_{\mbox{\textsc{co}}}$ and its dependence on galaxy parameters, finding a weak dependence of $X_{\mbox{\textsc{co}}}$ on temperature from 20 K to 100 K, and also finding $X_{\mbox{\textsc{co}}}$ increases for surface densities > 100 M$_\odot$ pc$^{-2}$ due to saturation of the CO line in the high-surface density regime.

It is important to better understand this relationship between metallicity and $X_{\mbox{\textsc{co}}}$, especially as we are now observing high-redshift low-metallicity galaxies from the early Universe using ALMA \citep{fujimoto_jwst_2024}. \citet{hu_co-evolution_2023} looked at simulating CO from a low metallicity (<0.1Z$_\odot$) galaxy to match the observation of the Wolf-Lundmark-Melotte (WLM) galaxy made by \citet{rubio_dense_2015}, and how dust evolution affects the models created. A hybrid method for the chemical evolution was adopted, where the H network was solved on-the-fly and the C chemistry was post processed using \textsc{astrochemistry}\footnote{\href{https://github.com/huchiayu/AstroChemistry.jl}{https://github.com/huchiayu/AstroChemistry.jl}.}. When including dust growth in their models, they were able to reproduce the observed CO luminosity. The conversion factor that was simulated was found to be close to the Milky Way value, although dust-based observations would suggest it should be higher due to the low metallicity. The authors conclude that using a metallicity-dependent $X_{\mbox{\textsc{co}}}$ factor would underestimate the dust-to-gas ratio in their simulations.

Other computational studies that have looked into modelling CO and the $X_{\mbox{\textsc{co}}}$ conversion factor include \citet{keating_reproducing_2020}, which used the cosmological zoom-in simulations from the FIRE project \citep{hopkins_fire-2_2018} and combined them with a post-processing method assuming chemical equilibrium using the \textsc{chimes} chemistry solver\footnote{\href{https://richings.bitbucket.io/chimes/home.html}{https://richings.bitbucket.io/chimes/home.html}} \citep{richings_non-equilibrium_2014-1,richings_non-equilibrium_2014}, allowing them to directly model the line emission from the simulations using the line radiative transfer code \textsc{radmc-\oldstylenums{3}d} \citep{dullemond_radmc-3d_2012}\footnote{\href{http://www.ita.uni-heidelberg.de/~dullemond/software/radmc-3d/}{http://www.ita.uni-heidelberg.de/~dullemond/software/radmc-3d/}.}. When comparing the column densities of H$_2$ and CO in these investigations, it was found that the shielding length in the simulations had to be reduced by one or two orders of magnitude for the simulated column densities to match with observational data. It was also noted that the $X_{\mbox{\textsc{co}}}$ calculated for their Milky Way-like galaxy was an order of magnitude lower than expected from the Milky Way value of 2 $\times$ 10$^{20}$cm$^{-2}$(K km s$^{-1})^{-1}$. This was noted to potentially be due to the resolution of the simulation, and that simulations with higher resolution could help fix this.

Higher resolution simulations, such as the SILCC simulations \citep{walch_silcc_2015}, have simulated pieces of galactic discs, about 500 pc $\times$ 500 pc $\times$ $\pm$5 kpc. These simulations however solve both the H$_2$ and CO chemistry on-the-fly, rather than post-processing. A subset of these simulations, SILCC-Zoom \citep{seifried_silcc-zoom_2017}, used a spatial resolution of 0.06 pc to study the formation and evolution of molecular clouds in high resolution. When examining the CO-to-H$_2$ conversion factor, they found the average $X_{\mbox{\textsc{co}}}$ scattered around 1 - 4 x 10$^{20}$cm$^{-2}$(K km s$^{-1})^{-1}$, agreeing with observational data and \citet{bolatto_co--h_2013}. It was also found that in these high resolution simulations, the average $X_{\mbox{\textsc{co}}}$ increased over time in agreement with \citet{glover_is_2016}. Other high resolution simulations include \citet{gong_environmental_2020}, which used the TIGRESS simulation framework \citep{kim_three-phase_2017}, simulating sections of a Milky Way-like galaxy at various radii from the galactic centre up to a resolution of 1 pc. These simulations were able to reproduce the relation between metallicity and $X_{\mbox{\textsc{co}}}$, finding that $X_{\mbox{\textsc{co}}}$ decreases with a power-law slope of -0.8 for the CO (1-0) line. This metallicity scaling of $X_{\mbox{\textsc{co}}}$ agrees with prior work by \citet{feldmann_x-factor_2012-1,feldmann_x-factor_2012}, which combined non-equilibrium chemistry for H$_2$ and full radiative transfer cosmological simulations with a subgrid CO model. 

\citet{richings_effects_2016,richings_chemical_2016} also found agreement with \citet{glover_is_2016}, finding that the average $X_{\mbox{\textsc{co}}}$ decreases with time within their low metallicity (0.1 Z$_\odot$) non-equilibrium simulations. However, at ages > 15 Myr, they note that this trend is uncertain, as there are too few clouds of a high age. Within \citet{richings_effects_2016}, they compared the effects of non-equilibrium chemistry on their simulations, and found that CO emission and H$_2$ mass can be both enhanced and suppressed by non-equilibrium chemistry, meaning that the average $X_{\mbox{\textsc{co}}}$ could be affected up to a factor of \textasciitilde 2.3. They also note that usage of a fluctuating UV radiation field would influence these non-equilibrium effects further and may drive additional effects not captured in their current simulations.

In this paper, we use a suite of isolated galaxy simulations that couple the CHIMES chemistry and cooling module to the FIRE-2 subgrid models for galaxy formation \citep{richings_effects_2022}. We also quantify the effects of non-equilibrium chemistry on those results compared to simulations where we assume chemical equilibrium.

 The remainder of this paper is organised as follows. In section \ref{Methods} we describe the methods we used to create and evolve isolated galaxy simulations, how we track the chemistry within, and how we create simulated spectra from the simulations. In section \ref{Simulations} we detail the isolated galaxy simulations we use, their initial conditions and evolution. In section \ref{Observations} we compare our simulations and simulated spectra to observational data such as xCOLD GASS and dust observations from the Milky Way. In section \ref{XCO_section} we look at the conversion factor and its relation to metallicity, as well as prescribing a new fitting for a multivariate function for the CO-to-H$_2$ conversion factor. We also test whether $X_{\mbox{\textsc{co}}}$ changes when averaged and blurred over different spatial scales to better match our simulated spectra with observations. In section \ref{Conclusions} we summarise our conclusions and our main findings. In Appendix \ref{ResTests} we show results from our resolution tests.

\section{Methods} \label{Methods}
Here we detail the methods used to create the simulations, and then from them the simulated spectra that we use throughout this work. We use the simulations from \citet{richings_effects_2022} that were run with their fiducial model. That work also considered model variations with a uniform radiation field, and using a constant dust-to-metals ratio with no dust depletion. However, for this study we use their fiducial model, with local stellar fluxes computed from star particles and a density and temperature dependent model for the dust depletion factors, which we describe in section \ref{ChimesSubSec} below.

\subsection{FIRE subgrid models} \label{FireSubModels}
The simulations within this work were run with a meshless finite mass solver, using the gravity and hydrodynamics code \textsc{gizmo} \citep{hopkins_new_2015}. These simulations were run using the \textsc{fire-2} sub-grid physics models \citep{hopkins_fire-2_2018} which were developed as part of the FIRE collaboration\footnote{\href{ https://fire.northwestern.edu/}{ https://fire.northwestern.edu/}}. 

Stars can form from gas particles that become self-gravitating, Jeans-unstable, and are above a density threshold $n$ > 10$^3$ cm$^{-3}$. Details of the star formation algorithm are further described in appendix C of \citet{hopkins_fire-2_2018}. We assume each star particle is a single stellar population, where we assume a \citet{kroupa_variation_2001} initial mass function. Stellar feedback from each star particle is then calculated with simple fits to the stellar population models from \textsc{starburst}\oldstylenums{99} \citep{leitherer_starburst99_1999}. The stellar feedback mechanisms include Type II and Ia SNe, stellar winds from OB and AGB stars, and radiative feedback. 

The FIRE-2 models track the enrichment and evolution of the 11 elements that are used in the \textsc{chimes} chemistry solver. Star particles inject metals through SNe and stellar mass loss, the yields we use are as follows: Type Ia SNe yields from \citet{iwamoto_nucleosynthesis_1999}, Type II SNe yields from \citet{nomoto_nucleosynthesis_2006}, and OB/AGB yields from \citet{van_den_hoek_new_1997}, \citet{marigo_chemical_2001}, and \citet{izzard_new_2004}, which are summarised in appendix A of \citet{hopkins_fire-2_2018}. Type Ia SNe feedback is calculated using \citet{mannucci_two_2006}, for both the prompt and delayed stellar populations. SNe and stellar winds use a feedback system described in appendix D of \citet{hopkins_how_2018} (see also \citet{hopkins_fire-2_2018}).

\subsection{The CHIMES chemistry module} \label{ChimesSubSec}
We use the \textsc{chimes} chemistry and cooling module to track the evolution of 157 molecules and ions \citep{richings_non-equilibrium_2014-1,richings_non-equilibrium_2014}. This includes the ionisation states of H, He, C, N, O, Ne, Mg, Si, S, Ca and Fe, as well as 20 molecular species; including H$_2$ and CO.

As \textsc{chimes} is capable of post-processing the abundances assuming chemical equilibrium whilst also solving the non-equilibrium abundances on-the-fly, we are able to make a direct comparison between the two (e.g., \citeauthor{richings_effects_2016} \citeyear{richings_effects_2016}; \citeauthor{richings_effects_2022} \citeyear{richings_effects_2022}). Calculating these non-equilibrium abundances on-the-fly adds a large computational cost to the simulations (5-10$\times$), so comparing non-equilibrium and equilibrium results allows us to determine what effect the non-equilibrium chemistry has on the simulation

A fluctuating time-dependent UV radiation field is used within this work, which follows the propagation of radiation from star particles using a method based on the \textsc{lebron} algorithm used to model stellar radiation pressure in \textsc{fire} \citep{hopkins_fire-2_2018,hopkins_radiative_2020}. As the simulations are coupled with \textsc{chimes}, this method is modified. The \textsc{lebron} method tracks three stellar fluxes from all particles; the IR, optical, and UV bands. However, for photochemistry, the radiation is tracked in eight separate stellar age bins allowing accurate tracking of the age-dependence of the UV spectra. The radiation from each age bin is then divided into further bands, splitting the non-ionising FUV band from the ionising EUV band, giving 16 stellar fluxes in total for the photochemistry. The optical and IR bands are not required for the photochemical reactions.

Shielding from this radiation field within local gas clouds is based on a Sobolev-like approximation, which uses a shielding length ($L_{\mbox{sh}}$) calculated from the density gradient $\nabla \rho$ and the mean inter-particle spacing, $h_{\mbox{inter}}$, like so:

\begin{equation} \label{SobolevEqn}
    L_{\mbox{sh}} = \frac{1}{2} \left(\frac{\rho}{\nabla\rho} + h_{\mbox{inter}}\right),
\end{equation}

where the first term of this equation accounts for the size of the gas cloud around the particle, and the second term accounts for the size of the particle. We include a factor of 1/2 in this equation so that it matches equation A10 of \citet{gnedin_environmental_2011} in the limit where the density term dominates. The local column density of species $i$ is then given by $N_i =$ $n_{i}L_{\mbox{sh}}$ where $n_i$ is the density of species $i$. Using the methods described in \citet{richings_non-equilibrium_2014}, the photochemical rates are then suppressed as a function of the local column densities of H\textsc{i}, H$_2$, He\textsc{i}, He\textsc{ii}, CO, and dust. This equation treats the shielding using an average shielding length, and therefore column density for the local gas cloud. Therefore, this treatment can tend to overestimate shielding, which can lead to higher molecular abundances than expected. 

The \textsc{chimes} non-equilibrium chemistry solver also requires the gas-phase abundances of each element in the reaction network to be input. The simulations track the total elemental abundances, but for some elements, a fraction of their total elemental abundance will be locked up in dust grains and therefore will not contribute to the gas-phase chemistry. For that reason, we need to determine the fraction of each element in dust grains, and then reduce the gas-phase abundances to match. 

We use the density and temperature-dependent dust depletion model developed by \citet{richings_effects_2022}. This model uses observations from \citet{jenkins_unified_2009}, which determined the depletion fraction of metals onto dust grains within the solar neighbourhood, by measuring the column densities of 17 metals and H\textsc{i} along 243 sightlines in the Milky Way. The model also uses observations from \citet{de_cia_dust-depletion_2016}, which expanded on these observations by recording 70 damped Lyman-$\alpha$ absorbers observed through quasar spectra, in addition to the original Milky Way sightlines. \citet{jenkins_unified_2009} found that the overall strength of dust depletion, $F_\ast$, is best fit to the relation:

\begin{equation}
    F_\ast = 0.772 + 0.461\log_{10} \langle n_{\textsc{H}} \rangle.
\end{equation}

This relation can be used to calculate the strength of dust depletion for each gas particle as a function of density, assuming the particles total hydrogen density $n_{\textsc{H}, \mbox{tot}}$, is approximately equal to the average neutral hydrogen density along the line of sight to the background source in the observations, $\langle n_{\textsc{H}} \rangle$. However this assumption can overestimate the strength of the depletion. We also limit $F_\ast$ at higher densities ($n_{\textsc{H}, \mbox{tot}}$ > 3.12 cm$^{-3}$) which corresponds to the strongest dust depletion strength observed in the Milky Way sightlines, as anything higher would be extrapolating to higher depletions than are observed. A limit linked to temperature is also imposed, where above 10$^6$ K all metals are in the gas phase due to any dust grains being destroyed through sputtering \citep{tsai_interstellar_1995}. 

The depletion factors of individual elements are then linear fits as a function of the parameter $F_\ast$, with fit coefficients collated from \citet{jenkins_unified_2009} and \citet{de_cia_dust-depletion_2016}, though these fits are uncertain due to limited observations. \citet{jenkins_unified_2009} only contains a handful of carbon depletion measurements. \citet{sofia_determining_2011} finds the gas-phase column densities of carbon are a factor of $\approx$ 2 than those measured in \citet{jenkins_unified_2009}, which would result in a stronger depletion of carbon in dust than expected from the linear fits by that same factor of 2. These depletion factors are used to reduce the gas-phase abundance of each element. We also sum the mass in dust grains of the 17 elements featured in the two observational papers to determine the total dust abundance, and use that to scale the rate of reactions that occur on the surface of the dust grains and any thermal processes that involve dust grains e.g. photoelectric heating. 

As this model is based on Milky Way and damped Lyman-$\alpha$ absorbers, it does not explicitly follow the formation and destruction mechanics that govern the abundances of dust grains. However, other studies have been able to develop models that are able to capture these processes, allowing for a more detailed evolution of the dust grain population \citep{asano_dust_2013,bekki_dust-regulated_2015,hirashita_dust_2015,mckinnon_simulating_2018,choban_galactic_2022}.

\subsection{The RADMC-3D radiative transfer code} \label{radmc3d_subsec}
Once we have the final snapshots from the simulations (described in Section \ref{GalFormSub}), we post-process the CO (1-0) emission line using \textsc{radmc-3d} \citep{dullemond_radmc-3d_2012}. \textsc{radmc-3d} allows us to follow the emission, propagation and absorption of the spectral line, along with stellar emission. We are also able to follow the absorption, scattering, and thermal emission from dust grains within the simulation. This means we are also able to calculate the total emission including the continuum, as well as the emission from solely the continuum. Obtaining the line emission is then as simple as subtracting the former from the latter.

\textsc{radmc-3d} first constructs an adaptive mesh refinement (AMR) grid. We refine the grid until each cell contains no more than 8 star particles and/or gas particles. The abundances of the ion and molecules, as well as their ion-weighted temperatures and velocities, are projected onto the AMR grid. The star particles are also projected onto the grid, and are smoothed, after being split into the eight stellar age bins mentioned in section \ref{ChimesSubSec}. We also include graphite and silicate grains in the \textsc{radmc-3d} simulated spectra, using the abundances of each at solar metallicity from the grain abundances from v13.01 of the \textsc{cloudy} photoionisation code \citep{mathis_size_1977,ferland_2013_2013}, and we scale these for each gas particle first by the total metallicity relative to solar for each gas particle, and then by the density and temperature-dependent dust to metals ratio predicted by our dust depletion model. The dust temperature in each cell of the AMR grid is then calculated using stellar radiation.

We then calculate the level populations of ions and molecules within each cell using the local velocity gradient (LVG) method \citep{castor_spectral_1970,goldreich_molecular_1974,shetty_modelling_2011}, as this approximates the effects of non-local thermodynamic equilibrium (non-LTE). This method calculates an optical depth for each cell using the gas velocities from neighboring cells, and finds an escape probability for each photon by assuming each photon will eventually be Doppler shifted away from the line centre. Gas can also be excited using this, depending on the temperature. We use atomic data and collisional excitation rates from \textsc{lamda}\footnote{\href{https://home.strw.leidenuniv.nl/~moldata/}{https://home.strw.leidenuniv.nl/~moldata/}} \citep{schoier_atomic_2005} and \textsc{chianti} \footnote{\href{https://www.chiantidatabase.org/}{https://www.chiantidatabase.org/}} \citep{dere_chianti_1997,landi_chiantiatomic_2013}, and we also account for the effects of the cosmic microwave background. The line emission from each cell is then calculated using the level populations.

\textsc{radmc} then produces a 3D data cube in position-position-velocity space, with velocities spanning $\pm$200 km s$^{-1}$ about the line centre of the emission. This data cube has a spectral resolution of 2 km s$^{-1}$ and a spatial resolution of 20 pc for each simulated spectra created.

\section{Simulations} \label{Simulations}
\subsection{Galaxy formation models} \label{GalFormSub}
We describe the initial conditions of the simulations in table \ref{InitConditions}, with a complete insight into these initial conditions available in table 3 of \citet{richings_effects_2022}. The simulations range from dwarf galaxy halo masses ($M_{200, \mbox{ crit}}$ = 10$^{10}$M$_\odot$) up to Milky Way-mass ($M_{200, \mbox{ crit}}$ = 10$^{12}$M$_\odot$) where $M_{200, \mbox{ crit}}$ is the total halo mass. The total stellar masses were calculated using the abundance matching model from \citet{moster_galactic_2013}, which has been modified according to \citet{sawala_bent_2015} so the model can account for the inefficiencies of galaxy formation at lower halo masses. We also use two variants of the m3e11 galaxy with a low and high gas fraction, so we can test the effect of gas fraction on our results.

\begin{table} 
\begin{minipage}{84mm}
\centering
\caption{Initial conditions of the simulations used within this work. A more complete insight into these is included in \citet{richings_effects_2022}. $M_{200, \mbox{ crit}}$ is the total halo mass in M$_\odot$ of the galaxy, which each model is named after, $f_{\mbox{gas}}$ is the gas fraction, and $Z_{\mbox{init}}$ is the initial metallicity compared to $Z_\odot$. $Z_{0,\mbox{tot}}$ is the total metallicity at time $t = $ 0 Myr to show the enrichment that occurs during the 300 Myr settling-in period, and $Z_{0,\mbox{gas}}$ is the gas-phase metallicity when accounting for dust-depletion using the model described in Section \ref{ChimesSubSec}. $Z_{0,\mbox{tot}}$ and $Z_{0,\mbox{gas}}$ are mass-weighted averaged over all gas particles in the simulation.}
\label{InitConditions}
\begin{tabular}{cccccc}
  \hline
  Name & $M_{200, \mbox{ crit}}$ & $f_{\mbox{gas}}$ & $Z_{\mbox{init}}$ & $Z_{0,\mbox{tot}}$ &  $Z_{0,\mbox{gas}}$\\
   & (M$_\odot$) & & ($Z_\odot$) & ($Z_\odot$) & ($Z_\odot$)\\
  \hline
  m1e10 & $10^{10}$ & 0.90 & 0.06 & 0.09 & 0.07\\
  m3e10 & $3\times10^{10}$ & 0.77 & 0.3 & 0.33 & 0.21\\
  m1e11 & $10^{11}$ & 0.49 & 0.8 & 0.92 & 0.58\\
  m3e11 & $3\times10^{11}$ & 0.30 & 1.1 & 1.51 & 0.94\\
  m3e11\_lowGas & $3\times10^{11}$ & 0.10 & 1.1 & 1.25 & 0.81 \\
  m3e11\_hiGas & $3\times10^{11}$ & 0.50 & 1.1 & 1.84 & 1.17\\
  m1e12 & $10^{12}$ & 0.19 & 1.2 & 1.88 & 1.18\\
\hline
\end{tabular}
\end{minipage}
\end{table}

Each simulation has a mass resolution of 400 M$_\odot$ for each gas and star particle. Using the cosmological density parameters for the total matter and baryonic content of the Universe, $\Omega_{\mbox{m}}$ and $\Omega_{\mbox{b}}$ respectively, we multiply the baryonic particle mass by ($\Omega_{\mbox{m}}$ - $\Omega_{\mbox{b}}$)/$\Omega_{\mbox{b}}$ to find and set the mass of dark matter particles to 1910 M$_\odot$. We apply a constant gravitational softening length of 2.8 pc for star particles, and 1.6 pc for dark matter. For gas particles, we use an adaptive gravitational softening length which we set equal to the mean inter-particle spacing, down to a minimum gas softening length of 0.08 pc. For example, at the density threshold for star formation, $n_{\textsc{H}}$ = 10$^3$ cm$^{-3}$, the gas softening length is set to 2.2 pc.

Resolution tests of each simulation were also ran to test how numerical resolution affects our results. Low-resolution versions of each simulation are included with a mass resolution of 3200 M$_\odot$, eight times lower resolution than our fiducial model. For high-resolution tests, only the m3e10 galaxy is included due to the extended runtime of the increased resolution combined with the non-equilibrium chemistry running alongside it, and this high-resolution m3e10 was run with 8 times higher mass resolution, at 50 M$_\odot$. We include results for these resolution tests in Appendix \ref{ResTests}.

The initial metallicities\footnote{We use the solar elemental abundances from Table 1 of \citet{wiersma_effect_2009} throughout this work, where the total solar metallicity is Z$_\odot$ = 0.0129} mentioned in table \ref{InitConditions} were set according to the mass-metallicity relation of SDSS galaxies from \citet{andrews_mass-metallicity_2013}. As these simulations include the injection of metals from winds and SNe (described in Section \ref{FireSubModels}), this metallicity will evolve over time. The initial relative abundances between different metals were assumed to be solar, and the initial He abundance was scaled to be between primordial and solar, according to the total metallicity. 

The galaxies are evolved for 800 Myr from the initial conditions. The simulation were first run for a period of 300 Myr with the supernova time-scales drastically reduced, to enable the disk to settle into a quasi-steady state. Within this 300 Myr period, we see an increase in metallicity within our simulations as we can see in Table \ref{InitConditions} for both our $Z_{0,\mbox{tot}}$ and $Z_{0,\mbox{gas}}$. This is likely due to a sudden burst of star formation within this period, causing the simulation to be injected with metals.

We then used the snapshot at the end of this 300 Myr settling-in period as $t = 0$ Myr and ran the simulation for a further 500 Myr with the standard models, taking one snapshot every 100 Myr for a total of 5 snapshots per simulation. It is important to mention we do not use the data from the initial 300 Myr period within our analysis or in creating our \textsc{radmc} spectra.

\begin{figure*}
\subfloat[]{{\includegraphics*[trim={5.1cm 0 5.1cm 0},width=2\columnwidth]{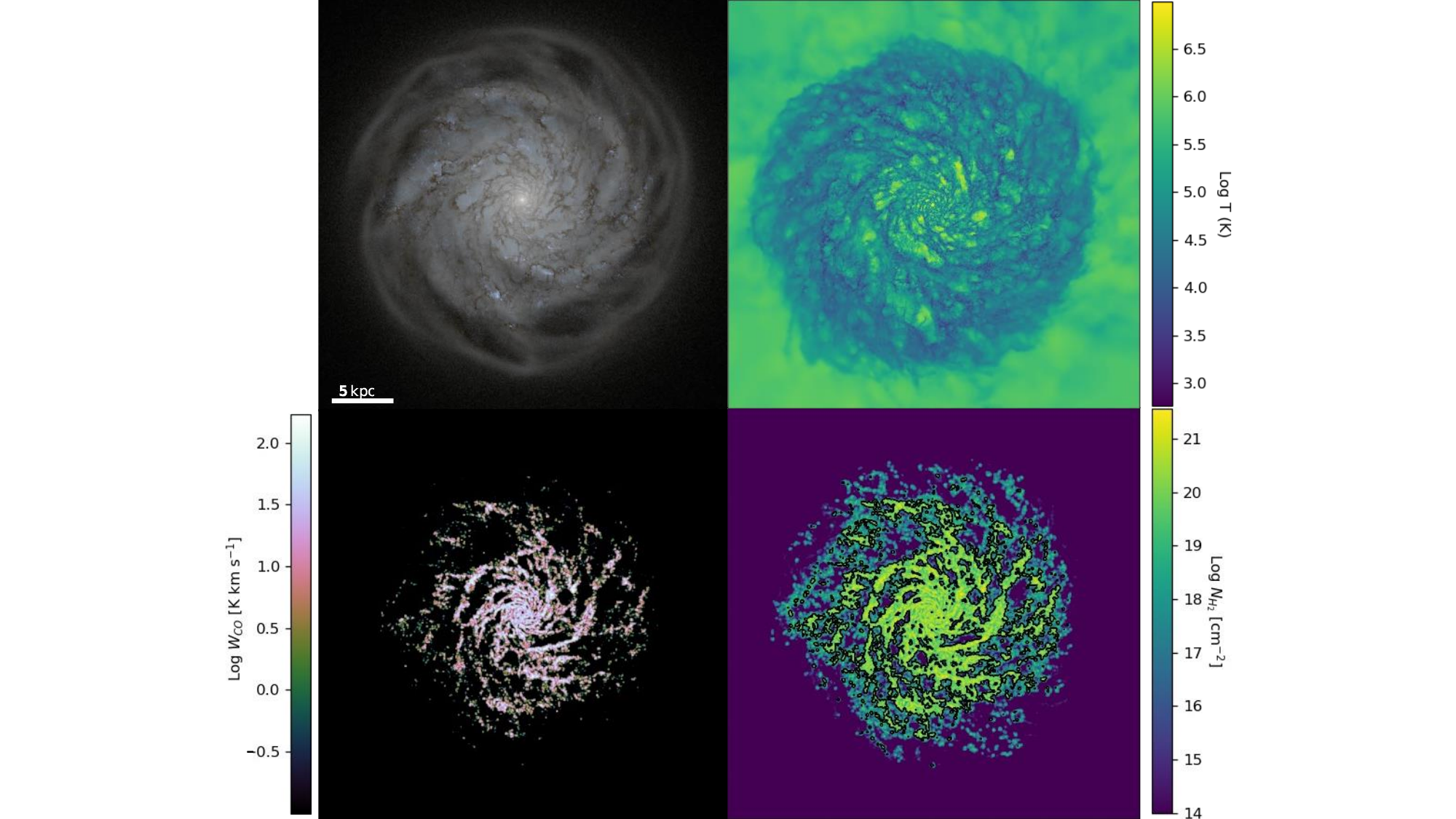} }}
\centering
\caption{Galaxy model m1e12 at time $t=$ 500 Myr shown in various ways
\textbf{Top left:} using a mock Hubble image created using post-processing the simulations through FIRE Studio. \textbf{Top right:} The mass-weighted temperature distribution of the gas. \textbf{Bottom left:} the velocity-integrated surface brightness of the CO emission made using \textsc{radmc}. \textbf{Bottom right:} Trace of the CO emission shown compared to the H$_2$ column density, showing the CO-dark gas. This trace is a contour showing all regions within that have velocity-integrated surface brightness > 0.1 K km s$^{-1}$.}
\label{m1e12images}

\end{figure*}

We show the Milky Way-mass galaxy fiducial simulation (m1e12) at 500 Myr in Figure \ref{m1e12images}, with a Hubble-like image made using the post-processing software FIRE Studio \citep{gurvich_fire_2022} in the left panel, and the mass-weighted gas temperature distribution in the right panel.

\subsection{Simulated emission spectra}
We run each simulation snapshot through the \textsc{radmc-3d} radiative transfer code mentioned in section \ref{radmc3d_subsec}, and we also run the 500 Myr snapshot with the post-processed abundances assuming chemical equilibrium. The final \textsc{radmc-3d} output is given in erg s$^{-1}$ cm$^{-2}$ Hz$^{-1}$ sr$^{-1}$, which we can convert into K km s$^{-1}$ using the equation below:

\begin{equation} \label{radio_to_wco_eqn}
    \frac{W_{\hbox{\textsc{co}}}}{\mbox{K km s$^{-1}$}} = \frac{1}{2} \frac{10^{-5}}{\mbox{km cm$^{-1}$}} \left({\frac{k_B}{\mbox{erg K$^{-1}$}}}\right)^{-1} \left(\frac{\lambda}{\mbox{cm}}\right)^3 \frac{\int I_\nu d\nu}{\mbox{erg s$^{-1}$ cm$^{-2}$ sr$^{-1}$}},
\end{equation}

where $k_B$ is the Boltzmann constant, $\lambda$ is the central wavelength of the emission, and $I_\nu$ is the line intensity which is given for each frequency $\nu$ in the range calculated by \textsc{radmc} from the central line. We can perform the integral on the right for each position within the 3D data cube we mentioned in section \ref{radmc3d_subsec}. After multiplying by the other factors used in the equation, using $\lambda =$ 0.26 cm for the CO (1-0) line emission, allows us to calculate and create a map of the velocity-integrated surface brightness W$_{\hbox{\textsc{co}}}$ like the one we show in Figure \ref{m1e12images}.

As can be seen in Figure \ref{m1e12images}, the CO emission follows the spiral structure of the galaxy, as well as within clusters around the arms. For this image, we filter out any surface brightness below 0.1 K km s$^{-1}$, but we do not do this for our main analysis and results unless we state otherwise.   


As we mentioned in section \ref{Introduction}, a portion of the H$_2$ gas within galaxies is CO-dark and cannot be traced by its emission. Using our simulated emission map, and the simulation data, we can create a trace of this emission against the H$_2$ column density to show how well our emission traces the gas with CO emission above the 0.1 K km s$^{-1}$ observational limit. We can also use this to find where gas that would be CO-dark to observers would be in our simulations, and what percentage of each simulation is CO-dark. This fraction of CO-dark gas expected to increase in low-metallicity environments \citep{amorin_molecular_2016,schruba_physical_2017,madden_tracing_2020}, which leads to an underprediction of the H$_2$ mass when using a constant conversion factor within low-metallicity galaxies. In Figure \ref{m1e12images}, we show the H$_2$ column density plot of our Milky Way-mass galaxy m1e12 with the simulated emission we see in Figure \ref{m1e12images} traced atop in black outline. This trace is a contour, showing regions within that have velocity-integrated surface brightness > 0.1 K km s$^{-1}$.

In Table \ref{COdarkTable} we show the percentage of CO-dark H$_2$ with $W_{\hbox{\textsc{co}}}$< 0.1 K km s$^{-1}$,$W_{\hbox{\textsc{co}}}$< 0.25 K km s$^{-1}$, and $W_{\hbox{\textsc{co}}}$< 1 K km s$^{-1}$, for each of our galaxy models in their 500 Myr snapshot for the fiducial model with non-equilibrium chemistry, and the model where we assume chemical equilibrium and post-process the abundances. These intensity cuts are based on observations from \citet{leroy_co--h_2011} of the Small Magellanic Cloud, which found the SMC to have a 3$\sigma$ intensity threshold of 0.25 K km s$^{-1}$, and NGC 6822 to have an intensity threshold of 0.03 K km s$^{-1}$.

\begin{table*} 
\begin{minipage}{84mm}
\noindent\makebox[\textwidth]{
\begin{tabular}{|l|l|l|l|l|l|l|}
  \hline
  \multirow{2}{*}{Name} & \multicolumn{2}{c}{< 0.1 K km s$^{-1}$} & \multicolumn{2}{c}{< 0.25 K km s$^{-1}$} & \multicolumn{2}{c|}{< 1 K km s$^{-1}$} \\
  &  non-Eqm & Eqm &  non-Eqm & Eqm &  non-Eqm & Eqm\\ 
  \hline
  m1e10 & 97.6\% & 95.0\% & 99.8\% & 99.9\% & 100\% & 100\%\\ 
  m3e10 & 94.4\% & 88.4\% & 94.8\% & 91.3\% & 98.2\% & 96.3\%\\ 
  m1e11 & 71.7\% & 77.2\% & 87.7\% & 84.5\% & 93.1\% & 90.7\%\\ 
  m3e11 & 54.0\% & 53.5\% & 61.5\% & 55.5\% & 70.7\% & 64.2\%\\ 
  m3e11\_lowGas & 59.3\% & 59.3\% & 69.4\% & 64.2\% & 79.3\% & 74.4\%\\
  m3e11\_hiGas & 41.5\% & 40.9\% & 51.5\% & 45.2\% & 60.5\% & 53.8\%\\
  m1e12 & 40.4\% & 34.3\% & 45.3\% & 38.5\% & 53.5\% & 46.3\%\\ 
\hline
\end{tabular}}
\caption{The percentage of H$_2$ gas which is considered CO-dark for each intensity cut ($W_{\mbox{\textsc{co}}}$ < 0.1 K km s$^{-1}$, $W_{\mbox{\textsc{co}}}$ < 0.25 K km s$^{-1}$, $W_{\mbox{\textsc{co}}}$ < 1 K km s$^{-1}$) within each galaxy at 500 Myr for the non-equilibrium (non-Eqm) and equilibrium (Eqm) simulations.} \label{COdarkTable}
\end{minipage}
\end{table*} 

Recent investigations using the SILCC-zoom simulations found the fractions of CO-dark H$_2$ gas varying from 15\% to 65\% when applying an observational detection limit of 0.1 K km s$^{-1}$, which increased to 20\% to 75\% when this detection limit was set to 1 K km s$^{-1}$ \citep{seifried_silcc-zoom_2020}. These simulations were in agreement with previous studies by \citet{levrier_uv-driven_2012}, which saw that the densest parts of the molecular gas in their simulations were 35\% to 40\% CO-dark. Another study we can compare to is \citet{richings_chemical_2016}, which found CO-dark fractions of 76\%-86\%, in their models with a halo mass matching our m1e11 model. These fractions are not too dissimilar from our calculations for our m1e11 model, although in their work, they used an observational limit of 0.25 K km s$^{-1}$, which would explain the slightly higher fraction of CO-dark gas.

Observational limits for the percentage of CO-dark gas tend to vary, with gamma ray emissions showing more than 30\% of H$_2$ is CO-dark \citep{grenier_unveiling_2005}, and sub-pc resolution observations of OH in the boundary of the Taurus molecular cloud showing the CO-dark gas fractions range from 20\% to 80\% \citep{xu_evolution_2016}. As can be seen in Table \ref{COdarkTable}, our results are in good agreement with both observational data and prior computational work. We also see how the fraction of CO-dark gas decreases as metallicity and gas fraction increase within both our non-equilibrium and equilibrium models, as expected from observational studies \citep{amorin_molecular_2016,schruba_physical_2017,madden_tracing_2020}.




\section{Comparison to observations} \label{Observations}
\subsection{CO to H2 column density relation} \label{ObsCols}

\begin{figure*}%
    \centering
    \subfloat[Results with non-equilibrium chemistry]{{\includegraphics*[width=\columnwidth]{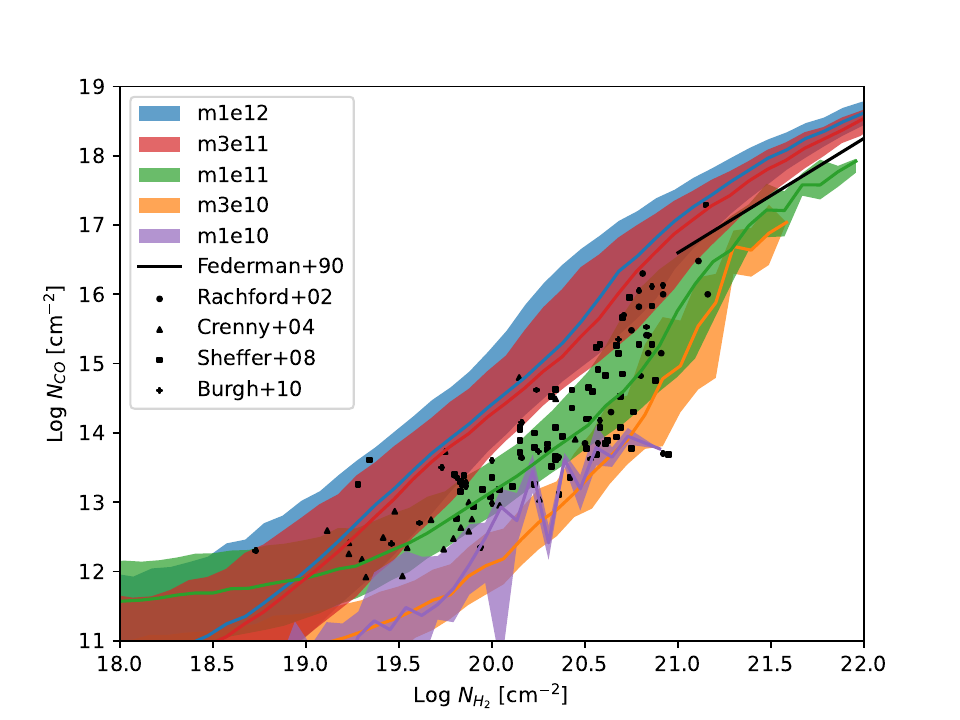} }}%
    \qquad
    \subfloat[Results with equilibrium chemistry]{{\includegraphics*[width=\columnwidth]{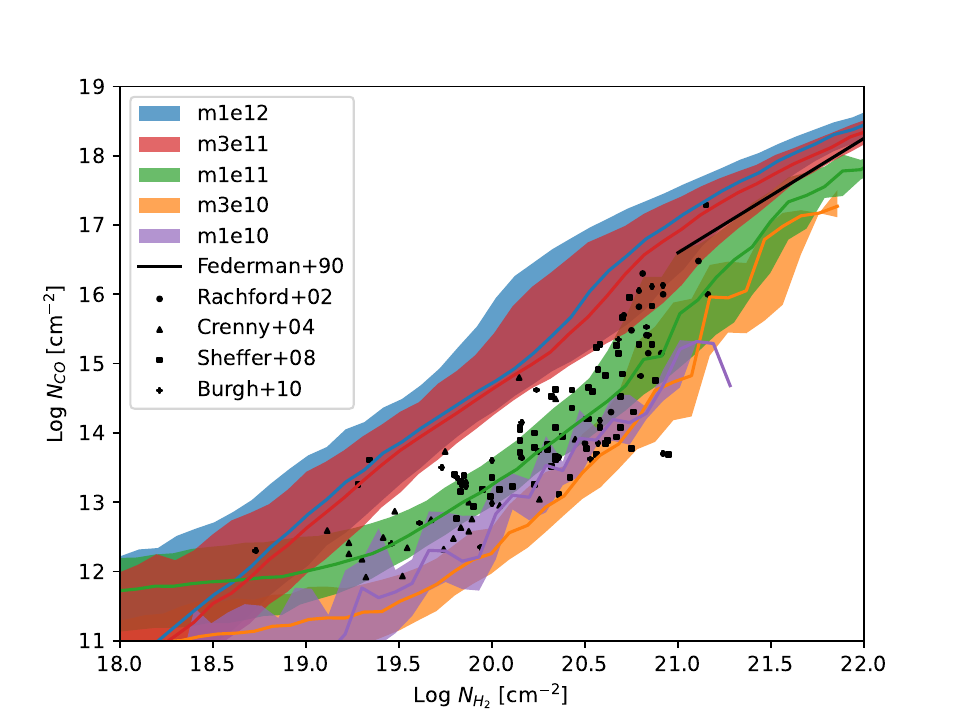} }}%
    \caption{Relation between CO and H$_2$ column densities for the range of halo masses. The shaded coloured regions indicate the 15th to 85th percentile within each H$_2$ column density bin, while the solid curves show the median, for each halo mass at time $t =$ 500 Myr. The black points included show observational data from UV absorption lines in the Milky Way from  \citet{federman_modeling_1990,rachford_far_2002,crenny_reanalysis_2004,sheffer_ultraviolet_2008}, and \citet{burgh_atomic_2010}.} \label{gasCols}
\end{figure*}

In this section we compare our simulation predictions for the relation between the column densities of CO and H$_2$ to observations compiled from UV absorption lines within the Milky Way. These observational surveys include \citet{federman_modeling_1990,rachford_far_2002,crenny_reanalysis_2004,sheffer_ultraviolet_2008}, and \citet{burgh_atomic_2010}. We show the readings from these observational studies, as well as data from the simulations we use in this work within Figure \ref{gasCols}. The shaded regions indicate the 15th to 85th percentile within each H$_2$ column density bin, while the solid curve shows the median, taken from time $t=500$ Myr within each simulation. The left- and right- hand panels show our simulation results using non-equilibrium and equilibrium abundances, respectively.

Previous work in \citet{keating_reproducing_2020} compared the observed relation between these Milky Way column densities to their simulations, and found their simulations over-produced CO by an order of magnitude at fixed H$_2$ column densities. To correct this, the shielding length needed to be decreased by a factor of 10 for the column densities to match the observed ratios. The authors suggest that the original discrepancy may have been due to the limited resolution of their simulations. The simulations in \citet{keating_reproducing_2020} also post-processed their chemical abundances, assuming chemical equilibrium, so we will see if our combination of higher resolution and non-equilibrium chemistry is able to remove this discrepancy.

As can be seen in Figure \ref{gasCols}, our non-equilibrium simulations cover the range of CO column densities at fixed H$_2$ column density from the observational studies. We see how the normalisation of this relation increases as the galaxy halo mass increases. This is due to the increasing metallicity of the simulations as the halo mass increases, which can be seen in prior work (see \citeauthor{hu__dependence_2022} \citeyear{hu__dependence_2022}, Figure 5). As the observations are from the Milky Way, we need to carefully consider which simulations are most appropriate to compare to. The m1e12 simulation has a similar halo mass to the Milky Way, but overproduces CO at a fixed H$_2$ column density by $\approx$0.5dex compared to the Milky Way observations. This is likely due to the higher metallicity in m1e12 as can be seen in Table \ref{InitConditions}. However, the total metallicity of our simulation m1e11 is close to solar metallicity, and reproduces the observed relation between CO and H$_2$ column densities seen in the Milky Way.  We therefore conclude that no further re-scaling of the shielding length (as in \citeauthor{keating_reproducing_2020} \citeyear{keating_reproducing_2020}) is necessary, and any remaining discrepancies between m1e12 and the Milky Way observations are primarily driven by the metallicity.

In our equilibrium simulations, we see that we predict a higher ratio of CO when compared to H$_2$. We therefore conclude that the improved agreement between the observations and our predicted CO and H$_2$ column densities, without the need for re-scaling of the shielding length, is due to a combination of our high resolution and the inclusion of non-equilibrium chemistry. 

Within Appendix \ref{ResTests}, we show how this column density relation is affected by resolution for our dwarf galaxy m3e10, and see how whilst we match the low-resolution at low column densities, at higher column densities we see that our fiducial model agrees more with our high-resolution version over the low-resolution.

We also compared the $N_{\mbox{\textsc{co}}}$ vs $N_{\mbox{\textsc{h}}_{\oldstylenums{2}}}$ relation in the high, fiducial, and low gas fraction variants of m3e11 (not shown) to see if the gas fraction of the simulation has any affect on the resulting ratio of CO and H$_2$ produced, and found no significant changes.

\begin{figure}%
    \centering
    \subfloat[m1e12 (Z = 1.9 Z$_\odot$)]{{\includegraphics*[trim={0cm 0cm 0cm 1.2cm},width=\columnwidth]{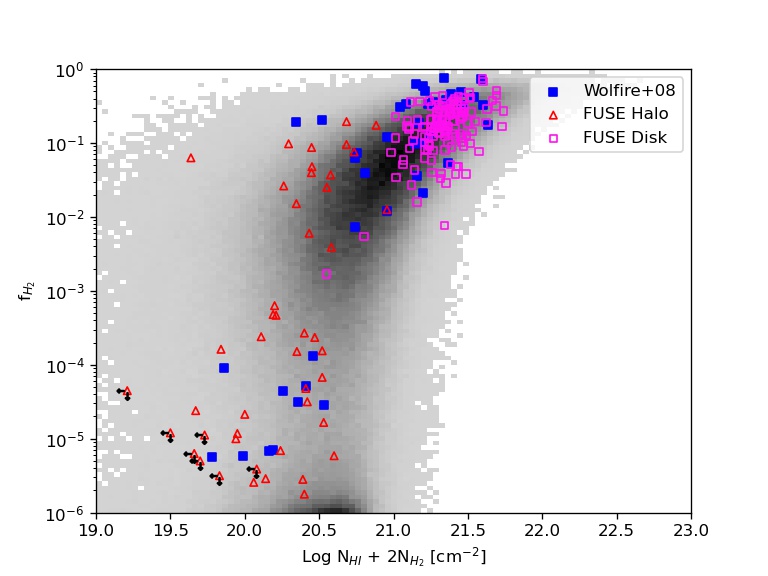} }}%
    \qquad
    \subfloat[m1e11 (Z = 0.9 Z$_\odot$)]{{\includegraphics*[trim={0cm 0cm 0cm 1.2cm},width=\columnwidth]{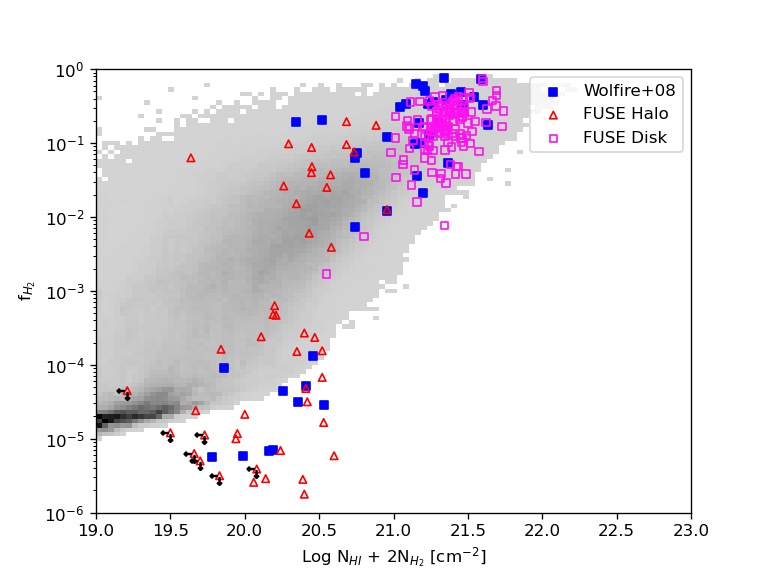} }}%
    \qquad
    \subfloat[m3e10 (Z = 0.3 Z$_\odot$)]{{\includegraphics*[trim={0cm 0cm 0cm 1.2cm},width=\columnwidth]{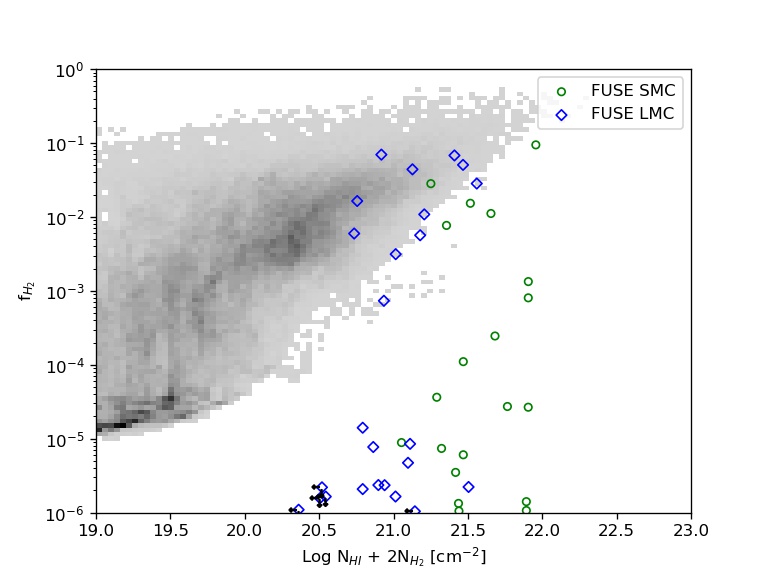} }}%
    \caption{The molecular gas fraction against the total neutral gas column density for \textbf{Top:} m1e12 at time $t =$ 500 Myr compared to measurements from the Milky Way. \textbf{Middle:} m1e11 at time $t =$ 500 Myr compared to measurements from the Milky Way. \textbf{Bottom:} m3e10 at time $t =$ 500 Myr compared to measurements from the LMC and SMC from FUSE. Milky Way observations collated from \citet{wolfire_chemical_2008,gillmon_fuse_2006} and \citet{shull_far_2021}. LMC and SMC observations from \citet{tumlinson_far_2002}. Observations which are upper-limits on the H$_2$ column density, and therefore the molecular gas fraction as well are shown with black arrows. The grayscale corresponds to the density of the plotted bins. We see our simulations are able to match the observations as well as the observed transitions from atomic-to-molecular gas.} \label{GasFracs}
\end{figure}

We further verify our H$_2$ model by comparing the fraction of molecular hydrogen, f$_{\mbox{\textsc{h}}_{\oldstylenums{2}}}$, against the total neutral gas column density using observations of H$_2$ absorption lines that have inferred the total neutral gas column density in Figure \ref{GasFracs}. We collate observations from surveys from the  Far Ultraviolet Spectroscopic Explorer (FUSE) of the galactic disk \citep{shull_far_2021}, the galactic halo \citep{gillmon_fuse_2006}, and the LMC and SMC \citep{tumlinson_far_2002}, as well as measurements of the Milky Way from \citet{wolfire_chemical_2008}. We group these observations and match them to the simulation that has the closest metallicity to them, similar to that done in \citet{gnedin_modeling_2009}. These are m1e11 at 0.9 Z$_\odot$ for the Milky Way observations, and m3e10 at 0.3 Z$_\odot$ for the LMC observations and SMC observations. We also include m1e12 and compare to the Milky Way observations as whilst m1e11 is closer in metallicity, m1e12 is similar in mass.

We see that whilst there are regions where our simulations match the observations, we also see some disagreement such as for m3e10 at low gas fractions (10$^{-6}$ to 10$^{-5}$). This could be due to how the simulations are measuring the column density compared to the observations, as in the simulations we are simply projecting the column densities onto an image grid, whereas the observations are based on UV absorption lines along a line of sight. This may account for the discrepancies we are seeing between the simulations and the observations. We do see however that our simulations line up with the observed trend of the transition from atomic-to-molecular gas decreasing with increasing metallicity \citep{gnedin_modeling_2009}.

\subsection{xCOLD GASS} \label{xCOLDSec} 
Another observational study we can use in comparing our simulations to is the xCOLD GASS survey \citep{saintonge_xcold_2017}, which we discussed in section \ref{Introduction}. As mentioned before, this survey includes CO measurements from 532 galaxies using the IRAM 30m telescope, measuring the CO (1-0) emission from each galaxy. In this section we will compare our simulation predictions with the directly observed CO luminosity and the inferred H$_2$ mass from the xCOLD GASS survey.

To calculate the CO luminosity for our simulations, we use an altered version of equation \ref{radio_to_wco_eqn} which we use to convert the \textsc{radmc} output into K km s$^{-1}$ pc$^{2}$. The xCOLD GASS survey calculates the CO luminosity using the observed frequency and luminosity distance \citep[see][Equation 2]{saintonge_xcold_2017}. This luminosity is then corrected to account for the aperture, and includes an error based on the measurement uncertainty, an 8\% flux calibration error, and a 15\% uncertainty due to the aperture correction. 

\begin{figure} 
    \includegraphics*[width=\columnwidth]{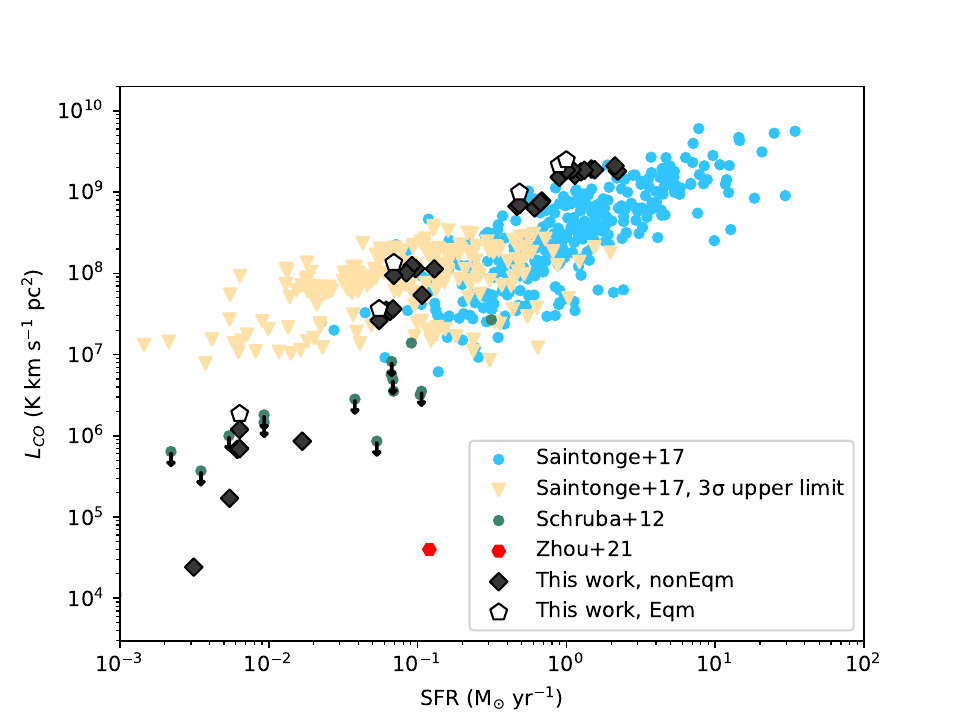}
    \caption{$L_{\mbox{\textsc{co}}}$ against SFR for each galaxy for each snapshot, with the non-equilibrium chemistry results in black diamonds, and the equilibrium chemistry results in white pentagons. The xCOLD GASS survey points are included in the blue, with the measurements of galaxies with no CO (1-0) emission detected shown in the yellow arrows, which are set as a 3$\sigma$ upper limit on the CO luminosity. We also include measurements from \citet{schruba_low_2012} in green and \citet{zhou_extremely_2021} in red to show readings in the low CO luminosity range to compare to our low-mass galaxies. Our predicted CO luminosities for our mid-to-high mass galaxies are able to match objects seen in the xCOLD GASS survey at fixed SFR, with our high-mass galaxies being \textasciitilde 0.5dex above the average luminosity for the xCOLD GASS survey at fixed SFR.} \label{LCO_SFR}
\end{figure}

We plot the 532 xCOLD GASS galaxies against our simulations using the CO luminosity that we calculate using \textsc{radmc} against star formation rate (SFR) in Figure \ref{LCO_SFR}. We show results for each 100 Myr snapshot for each galaxy simulation, and we include the results from our non-equilibrium chemistry simulations as well as our equilibrium simulations. Within our simulations, we calculate SFR by averaging the true mass of stars formed over the preceding 50 Myr in the simulations (see \citeauthor{floresvelazquez_time-scales_2021} \citeyear{floresvelazquez_time-scales_2021}). xCOLD GASS uses the sum of SFRs from the NUV and a WISE MIR band to probe unobscured and obscured star formation, respectively, which is described in \citet{janowiecki_xgass_2017}. We also include measurements from \citet{schruba_low_2012} and \citet{zhou_extremely_2021} to fill out the low CO luminosity/low SFR region to allow a comparison to our low-mass galaxies. Readings from \citet{schruba_low_2012} that are an upper-limit on the CO luminosity are depicted with an arrow indicating they should be lower, similar to those within the xCOLD GASS survey which we mentioned prior.

Whilst our mid-to-high gas-mass galaxies (with high SFR) are consistent with the most CO luminous galaxies seen in the xCOLD GASS observations at same SFR, they are nevertheless \textasciitilde0.5dex above the average relation seen in the observations. Given the limited sample size spanned by our suite of galaxy populations, it is unclear whether this is a systematic discrepancy or if our simulated galaxies are just extreme examples at fixed SFR. However, despite the high CO luminosities exhibited in our simulations, we have confirmed that our models produce the observed ratio of CO to H2 (based on their column densities, see Section \ref{ObsCols}), and therefore we are also producing an accompanying higher amount of H$_2$ in these simulations, meaning any calculation of the conversion factor using these galaxies should not be affected by the increased CO within.

Our low gas-mass galaxies, m1e10 and m3e10 lie below the cloud of xCOLD GASS observations. This is due to a lack of observational data in the xCOLD GASS survey at the mass ranges of these low-mass simulations. xCOLD GASS looks at galaxies with stellar masses of M$_\ast$ > $10^9M_\odot$, whereas our low-mass simulations of m1e10 and m3e10 have total stellar masses of 6.6 $\times$ $10^6$, and 8.9 $\times$ $10^7$ respectively. Therefore, these simulations fall far outside of the range of observations from the xCOLD GASS survey, so we would expect them to be below the data in our figure. On top of this, CO emission below 0.2$Z_\odot$ is near undetectable \citep{leroy_molecular_2005,schruba_low_2012,madden_tracing_2020}, meaning the metallicity range of these galaxies is incredibly hard to probe. xCOLD GASS contains many non-detections of CO at these star formation rates which they set as a 3$\sigma$ upper limit for the CO luminosity by assuming the full-width at half maximum of the H\textsc{i} emission is equal to the full-width at half maximum of the CO emission for a galaxy. Recent observations with ALMA have been able to detect CO emission at 0.1$Z_\odot$ in the WLM galaxy \citep{rubio_dense_2015}. We can see that measurements from \citet{schruba_low_2012} are measured at similar SFR as our m3e10 galaxy, with upper-limits on CO luminosity measurements placed above where we calculate the luminosity for our fiducial model to be, but slightly below where our equilibrium chemistry model is. More observations at these low-metallicity ranges will also help us better understand the nature of molecular gas in the low-metallicity regime by lowering these upper bound measurements from the xCOLD GASS survey, as well as those within \citet{schruba_low_2012}.

The chemical modelling in our simulations allows us to directly predict the H$_2$ mass. In contrast, the H$_2$ masses in xCOLD GASS are inferred from the CO luminosity, assuming the multivariate $\alpha_{\mbox{\textsc{co}}}$ conversion factor from \citet{accurso_deriving_2017}, which requires the metallicity and specific star formation rate of the observed galaxies. This Accurso function has a 35\% uncertainty within, and when combined with the error on the CO luminosity mentioned earlier gives a great uncertainty on this H$_2$ mass. On top of that, a portion of the xCOLD GASS galaxies were recorded with a non-detection of the CO (1-0) line, and therefore the values recorded for this are placed as a 3$\sigma$ upper limit on the H$_2$ gas mass reading. 

\begin{figure} 
    \includegraphics*[width=\columnwidth]{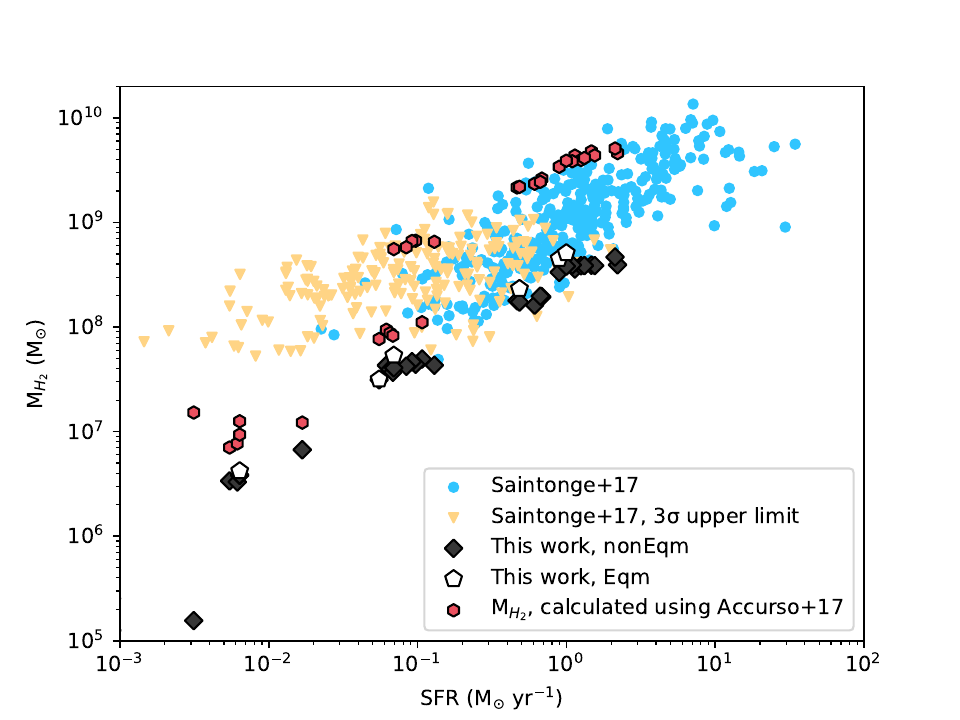}
    \caption{$M_{\mbox{\textsc{h}}_{\oldstylenums{2}}}$ against SFR for each galaxy for each snapshot with the non-equilibrium fiducial models in black, and the snapshots with equilibrium chemistry shown in white. Each galaxy mass groups by star formation rate, with m1e10 with the lowest SFR and m1e12 at the highest. The blue rings show xCOLD GASS survey points from galaxies where the CO (1-0) line emission was detected. Each calculated $M_{\mbox{\textsc{h}}_{\oldstylenums{2}}}$ has an error associated with it, which includes the error from the observed CO (1-0) luminosity and the error from the Accurso conversion function used to calculate $M_{\mbox{\textsc{h}}_{\oldstylenums{2}}}$. The yellow arrows are a 3$\sigma$ upper limit on the $M_{\mbox{\textsc{h}}_{\oldstylenums{2}}}$ from the galaxies with no CO (1-0) emission detected. The red hexagons show the $M_{\mbox{\textsc{h}}_{\oldstylenums{2}}}$ we calculate for each simulation using the Accurso multivariate function we show in equation \ref{alpCO_eqn}. We see that the Accurso function overpredicts the H$_2$ mass of our simulations by an order of magnitude.} \label{MH2SFR}
\end{figure}

We plot the H$_2$ mass from the xCOLD GASS galaxies and our non-equilibrium and equilibrium simulations against SFR in Figure \ref{MH2SFR}, and we highlight those galaxies with readings which are a 3$\sigma$ upper limit on the H$_2$ mass. We also plot the H$_2$ mass which we calculate using our simulated CO luminosity and the Accurso function we showed in equation \ref{Accurso_Eqn}. This is to show what an observer would calculate our simulated galaxies H$_2$ mass at, as the xCOLD GASS survey has done.

As before, our mid-to-high gas-mass galaxies m3e11, m3e11\_hiGas, and m1e12 are consistent with observed xCOLD GASS galaxies. As before, we see many non-detections with SFRs similar to our mid-mass galaxies. With CO (1-0) measurements of these galaxies which have their CO luminosities marked as a 3$\sigma$ upper limit by the xCOLD GASS survey, those observations will fall, though there is no lower constraints on the value of these galaxies yet meaning we do not yet know if they will lie within the bounds of our mid-gas mass simulations. However, our simulations line up with the lower edge of the envelope covered by the xCOLD GASS objects, when our calculated CO luminosities for those simulations were on the upper edge of the observations. We can also see our calculated H$_2$ mass using the \citet{accurso_deriving_2017} multivariate function fitting an order of magnitude higher than our actual H$_2$ mass for these simulations. Therefore if an observer was to measure our simulated galaxy and its CO luminosity, the calculated H$_2$ mass would be an order of magnitude too high if they used the \citet{accurso_deriving_2017} $\alpha_{\mbox{\textsc{co}}}$ factor. As this multivariate conversion factor was used within the xCOLD GASS survey, it is important that it can accurately estimate the H$_2$ mass of an observed galaxy. For these reasons, we discuss the Accurso function, and a possible replacement fitting further in section \ref{XCO_section}. At the bottom of these results once again is our low-mass galaxies m3e10 and m1e10. 

One possible reason for these discrepancies in $L_{\mbox{\textsc{co}}}$ and $M_{\mbox{\textsc{h}}_{\oldstylenums{2}}}$ could be the star formation model used within the simulations, as xCOLD GASS measures CO luminosities and calculates H$_2$ gas mass equal to what we find in the simulations. However, prior work by  \citet{orr_what_2018} verified the star formation model within the FIRE simulations, as they were able to reproduce a KS-like relation with a slope and scatter that matched observational data. On top of that, a change in the star formation we calculate would not make our results any closer to observations overall, as we require an increase in SFR to match the observations more closely for $L_{\mbox{\textsc{co}}}$, yet a decrease in SFR to match for $M_{\mbox{\textsc{h}}_{\oldstylenums{2}}}$. Therefore, we can rule out the star formation model as the source of the discrepancies. 

Within Appendix \ref{ResTests}, we analyse whether the numerical resolution has any effect on our simulated CO luminosity and H2 mass with our low and high resolution tests, and we compare these to the xCOLD GASS survey. We find similar results for our low resolution tests to our fiducial model, although the results for the high-resolution test of m3e10 is limited by the bursty nature of the dwarf galaxy models (see Appendix \ref{ResTests} for further discussion).



\section{Calibrating the CO-to-H2 conversion factor} \label{XCO_section}
In section \ref{Introduction} we discussed the CO-to-H$_2$ conversion factor, $X_{\mbox{\textsc{co}}}$, and how we can use it to estimate molecular gas content within galaxies using CO emission. Using our simulated CO emission, we are able to calibrate the conversion factor for each simulation as we have the exact H$_2$ mass available.



\subsection{Metallicity relation} \label{XCO_Met}
We discussed the relation between $X_{\mbox{\textsc{co}}}$ and metallicity earlier i.e. how when metallicity increases, the conversion factor decreases. \citet{bolatto_co--h_2013} gave an overview of dust-based observations, showing the decreasing $X_{\mbox{\textsc{co}}}$ with increased metallicity, as well as how $X_{\mbox{\textsc{co}}}$ rapidly increases below metallicities between 1/3 to 1/2 of solar. As we simulate galaxies across a range of metallicities, we are able to test this relation using our simulations. However, our low metallicity galaxies have extremely bursty star formation histories. This bursty star formation can cause the gas fraction of the disc to vary by a large factor \citep[see][Figure 7]{richings_effects_2022}, which can affect how much dense gas is within the galaxy, which is needed to form CO efficiently. 

\begin{figure} 
    \includegraphics*[width=\columnwidth]{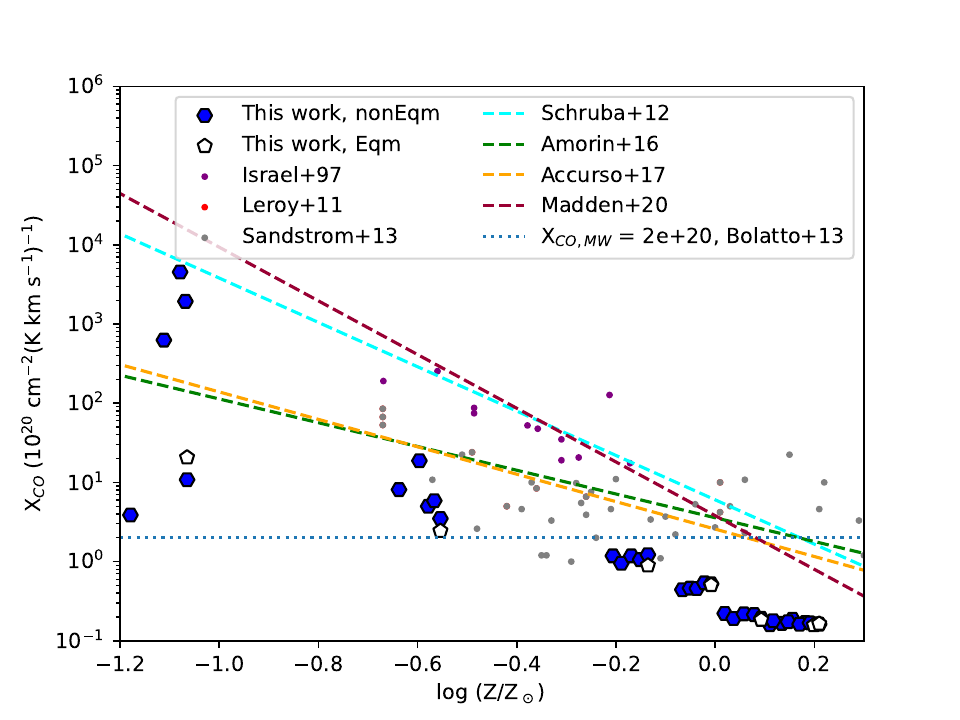}
    \caption{$X_{\mbox{\textsc{co}}}$ against gas-phase metallicity for each snapshot with the non-equilibrium fiducial models in blue, and the snapshots with equilibrium chemistry shown in white. We plot dust-based observational measurements from \citet{israel_h2_1997,leroy_co--h_2011} and \citet{sandstrom_co--h_2013}, as well as the $X_{\mbox{\textsc{co}}}$ for the Milky-Way with the blue-dotted line. We also plot metallicity-dependent relations for the conversion factor with coloured dashed lines from \citet{schruba_low_2012,amorin_molecular_2016,accurso_deriving_2017} and \citet{madden_tracing_2020}. We show metallicity when compared to solar metallicity, Z$_\odot$. Our simulations reproduce the observed trends between $X_{\mbox{\textsc{co}}}$ and metallicity for both our non-equilibrium and equilibrium models.} \label{XCO_Z}
\end{figure}

We compare to the dust-based approaches from \citet{israel_h2_1997,leroy_co--h_2011} and \citet{sandstrom_co--h_2013}. For each galaxy snapshot, we calculate the average $X_{\mbox{\textsc{co}}}$ across the galaxy and plot those in Figure \ref{XCO_Z} against the dust-based observations for both our non-equilibrium simulations, and those assuming chemical equilibrium. We also plot metallicity-dependent relations for the conversion factor with coloured dashed lines from \citet{schruba_low_2012,amorin_molecular_2016,accurso_deriving_2017} and \citet{madden_tracing_2020}.

We can see that both our non-equilibrium and equilibrium simulations follow the trend shown in the dust-based observations. For our low-metallicity galaxies like m1e10, which have highly bursty star formation, $X_{\mbox{\textsc{co}}}$ varies greatly from snapshot to snapshot. However, even with its bursty nature, the calculated $X_{\mbox{\textsc{co}}}$ for our low-metallicity galaxies are still in line with expectations of $X_{\mbox{\textsc{co}}}$ at low metallicities. 

\definecolor{pos_1_2}{rgb}{0.53186,0.66294,0.73784}
\definecolor{pos_1_3}{rgb}{0.6767,0.76722,0.81895}
\definecolor{pos_1_4}{rgb}{0.65102,0.74874,0.80457}
\definecolor{pos_1_5}{rgb}{0.6417,0.74202,0.79935}
\definecolor{pos_1_6}{rgb}{0.65171,0.74923,0.80496}
\definecolor{pos_1_7}{rgb}{0.72224,0.80001,0.84445}


\definecolor{pos_2_1}{rgb}{0.24229,0.45445,0.57568}
\definecolor{pos_2_2}{rgb}{0.92482,0.94587,0.9579}
\definecolor{pos_2_3}{rgb}{0.98852,0.99174,0.99357}
\definecolor{pos_2_4}{rgb}{0.71524,0.79497,0.84053}
\definecolor{pos_2_5}{rgb}{0.73694,0.8106,0.85269}
\definecolor{pos_2_6}{rgb}{0.70948,0.79083,0.83731}
\definecolor{pos_2_7}{rgb}{0.71878,0.79752,0.84251}


\definecolor{pos_3_1}{rgb}{0.24369,0.45546,0.57647}
\definecolor{pos_3_2}{rgb}{0.39387,0.22069,0.03}
\definecolor{pos_3_3}{rgb}{0.48307,0.33538,0.07692}
\definecolor{pos_3_4}{rgb}{0.94036,0.92332,0.8935}
\definecolor{pos_3_5}{rgb}{0.95355,0.94027,0.91705}
\definecolor{pos_3_6}{rgb}{0.9355,0.91707,0.88482}
\definecolor{pos_3_7}{rgb}{0.98908,0.99214,0.99389}

\begin{table}
\begin{minipage}{84mm}
\centering
\caption{Ratios of CO emission line luminosity, H$_2$ mass, and $X_{\mbox{\textsc{co}}}$, $Y_{\rm{noneq}} / Y_{\rm{eqm}}$, at 500~Myr using the fiducial model. Values highlighted in red and blue correspond to an enhancement and reduction, respectively, of these values when non-equilibrium abundances are used compared to post-processed equilibrium abundances.}
\label{noneq_vs_eqm_L}
\begin{tabular}{cccc}
\hline
& \multicolumn{3}{c}{$Y_{\rm{noneq}} / Y_{\rm{eqm}}$} \\ 
\cline{2-4} 
Galaxy & [$L_{\mbox{\textsc{co}}}$] & [$M_{\mbox{\textsc{h}}_{\oldstylenums{2}}}$] & [$X_{\mbox{\textsc{co}}}$] \\ 
\hline
m1e10 & 1 & \cellcolor{pos_2_1} \textcolor{white}{0.52376} & \cellcolor{pos_3_1} \textcolor{white}{0.52423}  \\
m3e10 & \cellcolor{pos_1_2} 0.6403 & \cellcolor{pos_2_2} 0.91725 & \cellcolor{pos_3_2} \textcolor{white}{1.43295}  \\
m1e11 & \cellcolor{pos_1_3} 0.72048 & \cellcolor{pos_2_3} 0.98642 & \cellcolor{pos_3_3} \textcolor{white}{1.36923} \\
m3e11 & \cellcolor{pos_1_4} 0.70483 & \cellcolor{pos_2_4} 0.74532 & \cellcolor{pos_3_4} 1.0426 \\
m3e11\_lowGas & \cellcolor{pos_1_5} 0.69932 & \cellcolor{pos_2_5} 0.76007 & \cellcolor{pos_3_5} 1.03318 \\
m3e11\_hiGas & \cellcolor{pos_1_6} 0.70524 & \cellcolor{pos_2_6} 0.7415 & \cellcolor{pos_3_6} 1.04607 \\
m1e12 & \cellcolor{pos_1_7} 0.75001 & \cellcolor{pos_2_7} 0.74768 & \cellcolor{pos_3_7} 0.98707 \\
\hline
\end{tabular}
\end{minipage}
\end{table}

We see in Figure \ref{XCO_Z} how our non-equilibrium and equilibrium results for $X_{\mbox{\textsc{co}}}$ appear similar. In Table \ref{noneq_vs_eqm_L}, we show the non-equilibrium effects on $L_{\mbox{\textsc{co}}}$ and $M_{\mbox{\textsc{h}}_{\oldstylenums{2}}}$ when compared to our equilibrium simulations. We then also quantify the non-equilibrium effects on $X_{\mbox{\textsc{co}}}$.

We see that non-equilibrium chemistry typically reduces $L_{\mbox{\textsc{co}}}$ and $M_{\mbox{\textsc{h}}_{\oldstylenums{2}}}$ by $\approx$ 70-76\%, atleast in the mid-to-high mass galaxies. However, as they are both reduced by a similar degree, the resulting $X_{\mbox{\textsc{co}}}$ factor is only weakly affected by a few percent. For our low-mass galaxies, we note that these simulations exhibit highly bursty star formation histories and gas fractions, so it is unclear whether the non-equilibrium effects suggested in Table \ref{noneq_vs_eqm_L} for low-mass galaxies will hold generally or may just be stochastic.

This result of seeing only a small fluctuation of $X_{\mbox{\textsc{co}}}$ between non-equilibrium to equilibrium contrasts previous studies by \citet{richings_effects_2016}, where the average $X_{\mbox{\textsc{co}}}$ within their simulations could be affected up to a factor of \textasciitilde 2.3. However, these simulations did not include a fluctuating UV radiation field, as well as link the UV field to the star particles, which we would expect to increase the results within non-equilibrium simulations. Our simulations also use different subgrid models, such as those for star formation and stellar feedback. It is possible the inclusion of one of these has driven additional effects, counteracting the non-equilibrium effects in the simulations, leaving only small fluctuations in the $X_{\mbox{\textsc{co}}}$ between non-equilibrium to equilibrium simulations for our mid-to-high mass galaxies. Further study is required to know which is causing the balance between $X_{\mbox{\textsc{co}}}$ in non-equilibrium and equilibrium simulations compared to previous studies.

\subsection{Multivariate conversion factor} \label{MultivariateSubsec}
Figure \ref{XCO_Z} demonstrates that $X_{\mbox{\textsc{co}}}$ is not a constant, and the conversion factor depends on conditions within the galaxy. We see that metallicity is one such factor that causes fluctuations in the conversion factor, but \citet{accurso_deriving_2017} demonstrated that other factors are also important by examining observed scaling relations between the ratio of luminosity of [C \textsc{ii}] to CO and local conditions from galaxies from the xCOLD GASS survey and the Dwarf Galaxy Survey (DGS) \citep{madden_overview_2013,remy-ruyer_gas--dust_2014,cormier_molecular_2014}. The factors that accounted for most of the variation in the luminosity ratio were metallicity, and offset from the main sequence, $\Delta$MS. Performing a linear fit to their radiative transfer models, they found the multivariate function we showed in equation \ref{Accurso_Eqn}. The version we show in equation \ref{Accurso_Eqn} uses the log of the ratio of the luminosity of [C \textsc{ii}] to CO, though there is a version without, instead using a constant in its place which we show below:

\begin{equation} \label{Accurso_Eqn2}
    \log \alpha_{\mbox{\textsc{co}}}(Z) = 15.623 - 1.732[12 + \log(\mbox{O/H})] + 0.051\log\Delta(\mbox{MS})\\
\end{equation}

\begin{figure*}%
    \centering
    \subfloat[Luminosity ratio against gas-phase metallicity]{{\includegraphics*[width=\columnwidth]{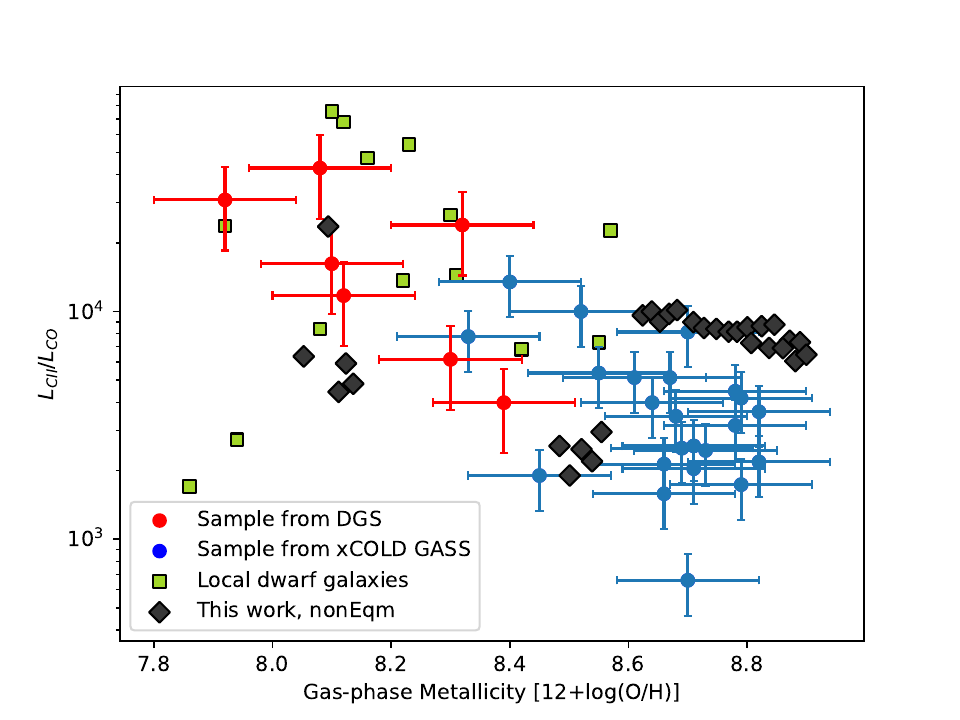} }}%
    \qquad
    \subfloat[Luminosity ratio against $\Delta$MS]{{\includegraphics*[width=\columnwidth]{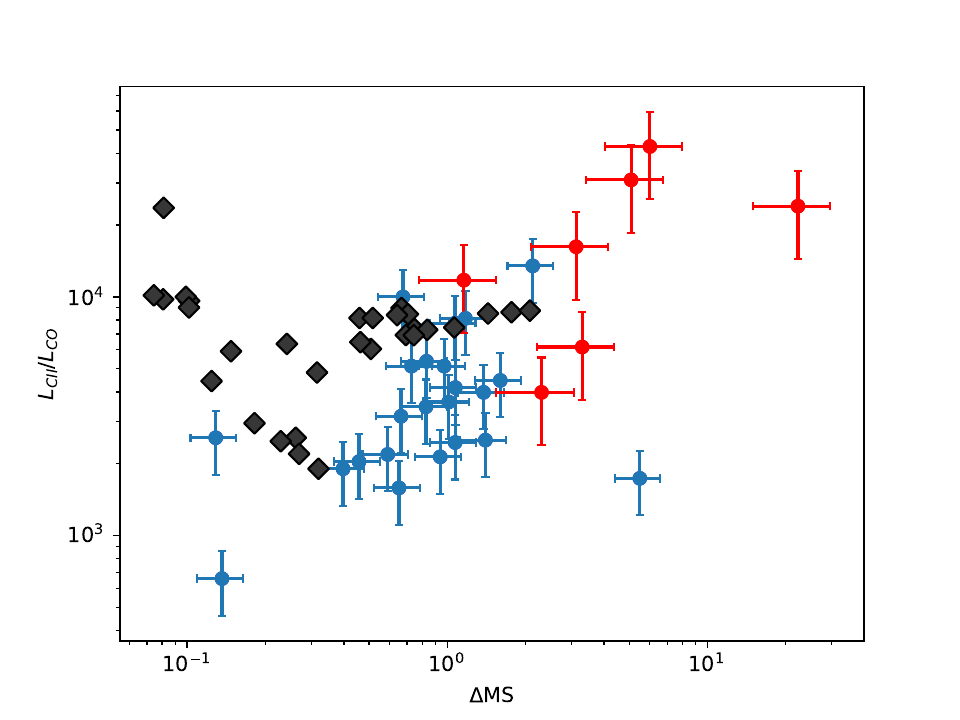} }}%
    \caption{Scaling relations of $L_{\mbox{\textsc{[C} {\textsc{ii}]}}}$/$L_{\mbox{\textsc{co}}}$ against gas-phase metallicity (left panel) and $\Delta$MS. Observations taken from a sample of 23 xCOLD GASS objects in blue, and 7 Dwarf galaxy survey (DGS) objects in red, as well as a sample of local dwarf galaxies from \citet{cormier_herschel_2015} in green. We plot our non-equilibrium fiducial models against in black. Whilst the simulations are broadly consistent with the observational data, the limited sample size in the simulations means they do not fully probe the complete distribution covered by the observations.} \label{FittingObs}
\end{figure*}

The substitution of the log of the luminosity ratio was achieved by performing a linear fit using those observed scaling relations to find an equation for the luminosity ratio in terms of metallicity and $\Delta$MS, and then substituting out the luminosity ratio. In Figure \ref{FittingObs} we show the observations used to find these scaling relations alongside our simulations. For the [C \textsc{ii}] luminosity, we use the results found in \citet{richings_effects_2022} to calculate the luminosity ratio with our $L_{\mbox{\textsc{co}}}$ calculated within this work. \citet{richings_effects_2022} found that non-equilibrium chemistry only affects their calculated ${L_{\mbox{\textsc{[C} {\textsc{ii}}\textsc{]}}}}$ by $\approx5$\%, and with our results in table \ref{noneq_vs_eqm_L} showing only a fluctuation of $\approx25$\% for the CO luminosity and 5 \% for the conversion factor, we only focus on our fiducial model results when calibrating our multivariate fitting.

The version of the function shown in equation \ref{Accurso_Eqn2} is used by the xCOLD GASS survey to calculate the H$_2$ mass for each galaxy they observe. As we showed in Figure \ref{MH2SFR}, this multivariate function overpredicts the mass of our galaxy models by up to an order of magnitude. For this reason, we use the method used in \citet{accurso_deriving_2017} to find a fitting from our simulations. 

Performing a linear fit to our simulations, we achieve a variation of the multivariate function which we show below:

\begin{equation} \label{new_Multivariate}
\begin{split}
    \log \alpha_{\mbox{\textsc{co}}}(Z) = a\log\frac{L_{\mbox{\textsc{[C} {\textsc{ii}}\textsc{]}}}}{L_{\mbox{\textsc{co}}(1-0)}} + b[12 + \log(\mbox{O/H})] \\
+ c\log\Delta(\mbox{MS}) + d,
\end{split}
\end{equation}
where $a = 2.8733 \times 10^{-2}$, $b = -2.0430$, $c = -9.7558 \times 10^{-3}$, and $d = 17.2542$. 

We also perform a fitting to match that of equation \ref{Accurso_Eqn2}, and find $a$ = 0, $b$ = -2.0362, $c$ = -0.0123, $d$ = 17.3036.

Using equation \ref{new_Multivariate} as a baseline with constants $a,b,c,$ and $d$, we show the constants for both versions of the Accurso function, and both versions of our fitting function in Table \ref{FittingConstTable}. We see for the luminosity ratio functions, our function sees more of a reliance on metallicity and less on $\Delta$MS. We also see that our function matches the observations with us seeing a negative correlation on metallicity, though we do not see the positive relation on $\Delta$MS due to not probing a full range of $\Delta$MS values within our simulations, which we can see in Figure \ref{FittingObs}. We see that when including results from \citet{cormier_herschel_2015}, our low-metallicity observations are within reason. We see that some simulations are also outside the region of the observations, which we will discuss shortly. The dependence of our fitting on the luminosity ratio is also reduced by a factor of \textasciitilde 30, with a greater dependence on the constant $d$. For both functions where $a = 0$, we see close agreement for $b$ and $d$, though even slight variations in these will cause large variations in the final $\alpha_{\mbox{\textsc{co}}}$.

\begin{table} 
\begin{minipage}{84mm}
\centering
\caption{Comparison of the constants $a$, $b$, $c$, and $d$ between both variations of the Accurso multivariate function, and both variations of the alternative multivariate function we propose in this work. The two functions where there is a non-zero constant $a$ are the functions where the ratio between luminosity of [C \textsc{ii}] to CO is used, and the two functions where $a$ is set to 0 are those where that ratio has been substituted out. The source column shows whether the equation comes from \citet{accurso_deriving_2017} (A+17) or this work (TW).} \label{FittingConstTable}
\begin{tabular}{ccccc}
  \hline
  Source & $a$ & $b$ & $c$ & $d$\\ 
  \hline
  A+17 & 0.742 & -0.944 & -0.109 & 6.439 \\
  A+17 & 0 & -1.732 & 0.051 & 15.623\\
  TW & 2.8733e-2  & -2.0430 & -9.7558e-3 & 17.2542 \\
  TW & 0 & -2.0362 & -0.0123 & 17.3036\\
\hline
\end{tabular}
\end{minipage}
\end{table}

\begin{figure} 
    \includegraphics*[width=\columnwidth]{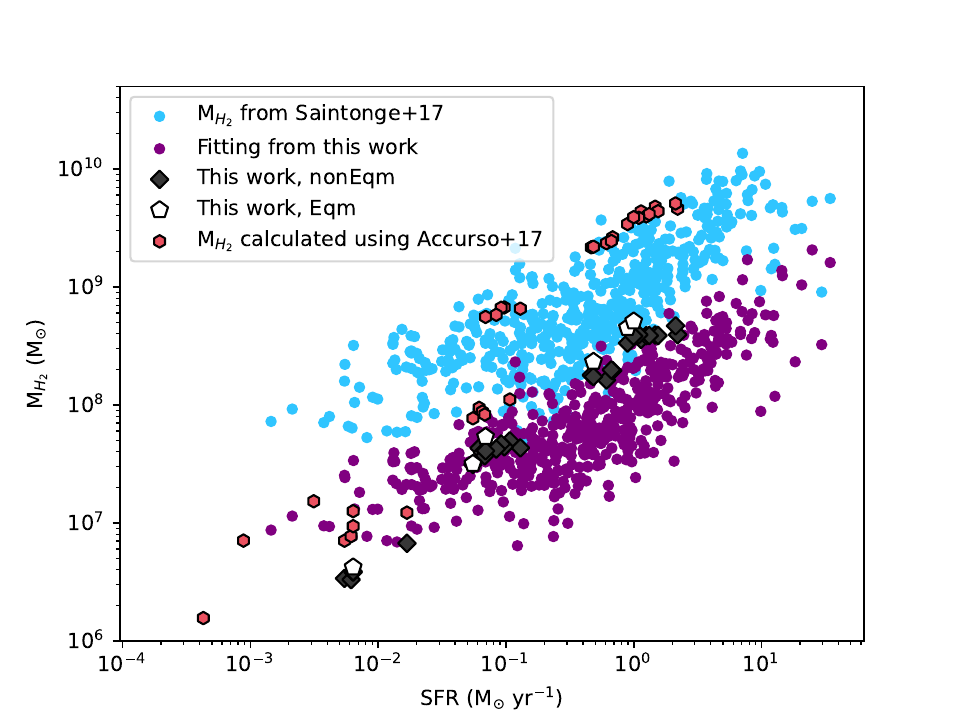}
    \caption{$M_{\mbox{\textsc{h}}_{\oldstylenums{2}}}$ against star formation rate. We show the $M_{\mbox{\textsc{h}}_{\oldstylenums{2}}}$ calculated within the xCOLD GASS survey using the Accurso multivariate function in blue, and compare to the $M_{\mbox{\textsc{h}}_{\oldstylenums{2}}}$ of the xCOLD GASS objects calculated using our alternative fitting function in purple. We also plot each of our snapshots with the non-equilibrium chemistry in black, the equilibrium chemistry in white, and the mass of our snapshots calculated using the Accurso function in red.} \label{Fitting_Fig}
\end{figure}

We plot the H$_2$ mass of the xCOLD GASS survey in Figure \ref{Fitting_Fig} against the H$_2$ mass calculated with our alternative fitting function with constants $a$ = 0, $b$ = -2.0362, $c$ = -0.0123, $d$ = 17.3036. We also plot both our non-equilibrium and equilibrium simulations, as well as the H$_2$ mass of our simulations calculated using the Accurso function. We can see our fitting estimates the H$_2$ mass by an order of magnitude below that which was calculated by the Accurso function for the xCOLD GASS survey. We also see our fitting now matches our $L_{\mbox{\textsc{co}}}$ predictions when compared to xCOLD GASS in Figure \ref{MH2SFR}, where our simulations were in the upper edge of the observations.

We caution that the fitting from our simulations is uncertain, as the limited range of galaxy conditions spanned by our simulation suite does not fully probe the complete distribution of metallicities and $\Delta$MS values observed in the xCOLD GASS survey and DGS, as seen in Figure \ref{FittingObs}. This may bias the results of these fits. With a greater range of simulations and more observational data, we would be able to see if our simulations truly match the relations between the luminosity ratio, metallicity, and $\Delta$MS. 



\subsection{Variation with spatial scale}

We have looked at how $X_{\mbox{\textsc{co}}}$ varies with factors such as metallicity or non-equilibrium chemistry effects, but one factor we also need to look at is spatial scale. Observations are not guaranteed to be as high-resolution as our simulations, so we need to test whether our simulated $X_{\mbox{\textsc{co}}}$ varies largely when adapting our simulated spectra to match observations more closely. We can investigate the variation of $X_{\mbox{\textsc{co}}}$ with different spatial scales through differing ways, the first is by using a Gaussian filter to blur the finer details of our simulations to better match observations. 

\begin{sidewaysfigure*}%
\centering
\includegraphics*[trim={3cm 0.5cm 3cm 2cm},width=\columnwidth]{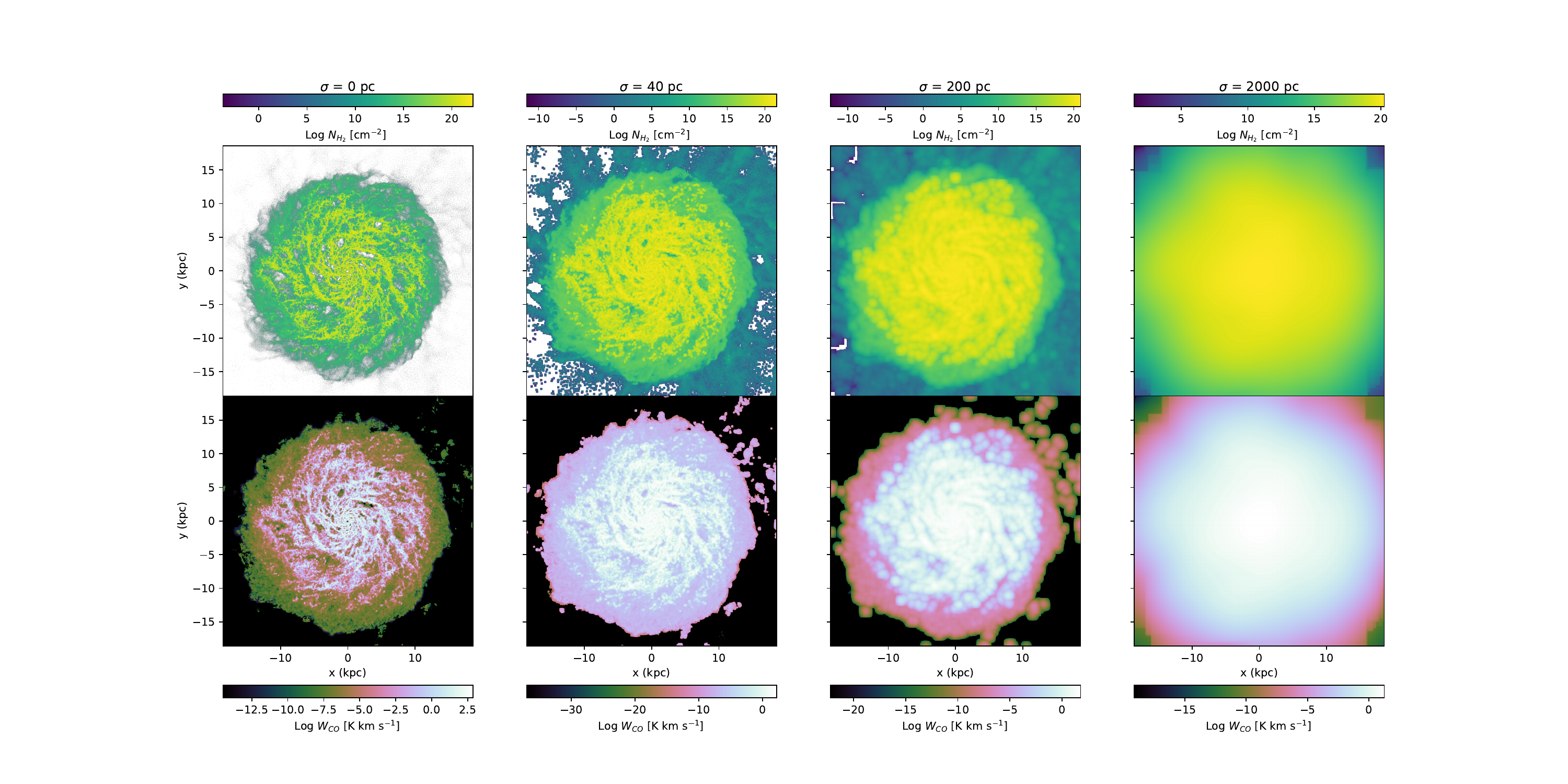}
\caption{Galaxy m1e12 at time $t = 500$ Myr, with a Gaussian filter applied, where the standard deviation of the filter, $\sigma$,  is increasing from left to right from 0 pc, 40 pc, 200 pc, up to 2000 pc. Each plot has their own respective colourbar above/below the figure. \textbf{Top:} The H$_2$ column density. \textbf{Bottom:} The velocity-integrated surface brightness of the CO (1-0) emission. As the $\sigma$ increases, the Gaussian filter removes all fine details for both the column density and the CO emission.} \label{GaussBlur_m1e12}
\end{sidewaysfigure*}

We show how this Gaussian filter affects the H$_2$ column density of our Milky-Way mass galaxy m1e12 as we increase the standard deviation of the filter, $\sigma$, in Figure \ref{GaussBlur_m1e12}. We also apply this filter to our velocity-integrated surface brightness $W_{\mbox{\textsc{co}}}$, which we see in the bottom half of Figure \ref{GaussBlur_m1e12}. We can see how the gas is smoothed out from $\sigma =$ 0 pc, with no blur applied, up to $\sigma =$ 2000 pc, where the galaxy is completely smoothed and all features have been lost. This Gaussian filter is in addition to the gas particles being smoothed over their kernel size with a cubic spline kernel.


\begin{figure} 
    \includegraphics*[width=\columnwidth]{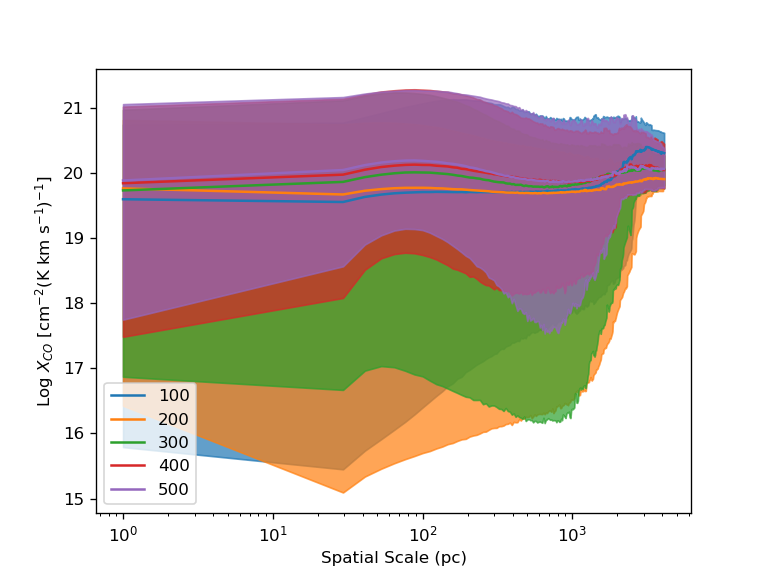}
    \caption{Relation between $X_{\mbox{\textsc{co}}}$ and the spatial scale of which the simulation has then been averaged over, which matches the FWHM of the Gaussian filter, for each snapshot taken every 100 Myr of galaxy m1e12 for our non-equilibrium fiducial model. The shaded region shows the 15th to 85th percentile of $X_{\mbox{\textsc{co}}}$ values, and the solid line shows the median, for each snapshot. We see how $X_{\mbox{\textsc{co}}}$ is near constant on all spatial scales, with some slight fluctuations.} \label{XCOGauss}
\end{figure}
We show how the average $X_{\mbox{\textsc{co}}}$ changes when the Gaussian filter `blur' is increased in Figure \ref{XCOGauss} for our Milky-Way mass galaxy m1e12 for each 100 Myr snapshot. We average the H$_2$ and $W_{\mbox{\textsc{co}}}$ on a spatial scale that matches the full-width at half-maximum (FWHM) of the Gaussian filter that has been applied. We can see that despite some fluctuations in the $X_{\mbox{\textsc{co}}}$ value, the average value for each snapshot increases by less than 0.5 dex from the initial average value at $\sigma = 0$ pc where no Gaussian blur is applied. The only exception to this is snapshot 100, which deviates by an order of magnitude when the Gaussian blur is at a FWHM of \textasciitilde3000 pc. The other deviations may be due to just slight variations in the H$_2$ gas distribution or $W_{\mbox{\textsc{co}}}$ causing much lower than expected values, which we can see in Figure \ref{GaussBlur_m1e12} where $\sigma = 40$ pc and $\sigma = 200$ pc, where the minimum values drop to much lower than without a blur by several orders of magnitude. We can see how the scatter on $X_{\mbox{\textsc{co}}}$ decreases on larger spatial scales as well (see also \citeauthor{feldmann_x-factor_2012} \citeyear{feldmann_x-factor_2012}). This is due to averaging on much larger spatial scales, meaning the resolution of our spectra decreases as $\sigma$ and the FWHM of our Gaussian filter increase. 
 

\subsection{Variation with other physical parameters}

In Section \ref{MultivariateSubsec}, we showed how the conversion factor is multivariate, but we only show metallicity and offset from main sequence as these values are easily available for an observer, and therefore our function is easily usable to calculate the conversion factor for observers. However, there are other parameters where we see variations in $X_{\mbox{\textsc{co}}}$, such as molecular gas surface density, $\Sigma_{\mbox{\textsc{h}}_{\oldstylenums{2}}}$.

Prior work by \citet{shetty_modelling_2011,shetty_modelling_2011-1} found that line saturation from lines of sight with high surface densities caused $X_{\mbox{\textsc{co}}}$ to rise. \citet{feldmann_how_2011} also found that on small scales, $X_{\mbox{\textsc{co}}}$ increases with $\Sigma_{\mbox{\textsc{h}}_{\oldstylenums{2}}}$. However, \citet{narayanan_co-h2_2011} found the opposite trend, and argued that as $\Sigma_{\mbox{\textsc{h}}_{\oldstylenums{2}}}$ increases, $X_{\mbox{\textsc{co}}}$ decreases due to an increase in velocity dispersion and kinetic temperature caused by more efficient dust-gas thermal exchange at higher densities. This increase in both the velocity dispersion and kinetic temperature caused an increase in CO intensity, and therefore a decrease in $X_{\mbox{\textsc{co}}}$.  This decrease occurs at the same range as the increase noted in \citet{shetty_modelling_2011,shetty_modelling_2011-1} (see Figure 6 in \citeauthor{narayanan_co-h2_2011} \citeyear{narayanan_co-h2_2011} and Figure 4 in \citeauthor{shetty_modelling_2011-1} \citeyear{shetty_modelling_2011-1}).

\begin{figure} 
    \includegraphics*[width=\columnwidth]{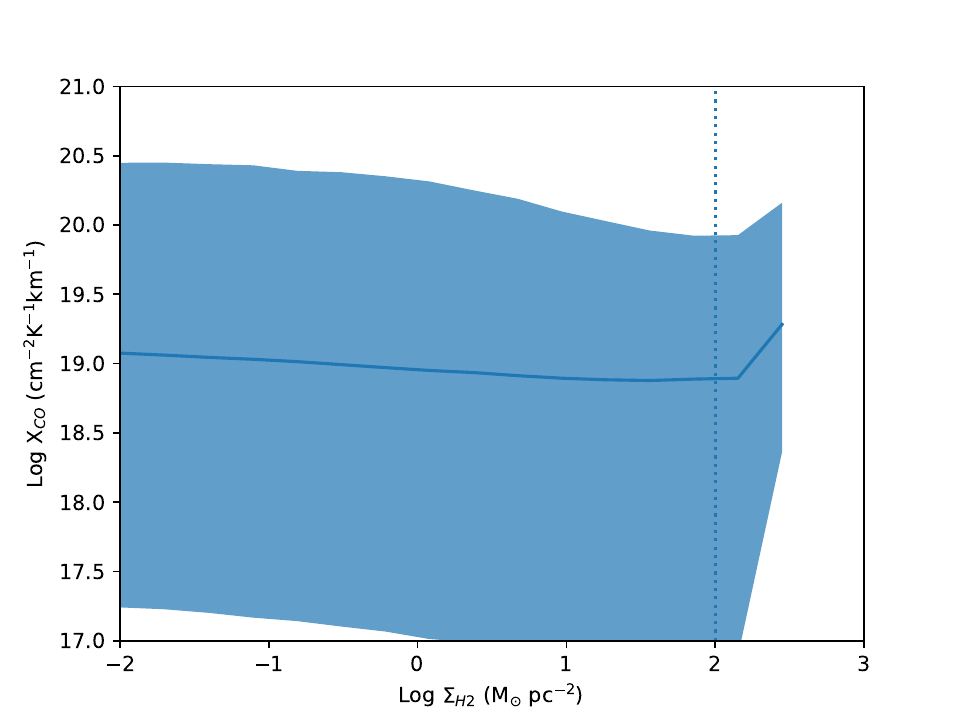}
    \caption{Relation between $X_{\mbox{\textsc{co}}}$ and $\Sigma_{\mbox{\textsc{h}}_{\oldstylenums{2}}}$ for galaxy m1e12 for our non-equilibrium fiducial model. The shaded region shows the 15th to 85th percentile of $X_{\mbox{\textsc{co}}}$, and the solid line shows the median. We also plot a dotted line at 10 $^2$ M$_\odot$ pc$^{-2}$ to show the upturn occurs at a similar range to \citet{shetty_modelling_2011-1}.} \label{SurfaceDensPlot}
\end{figure}

We show $X_{\mbox{\textsc{co}}}$ against $\Sigma_{\mbox{\textsc{h}}_{\oldstylenums{2}}}$ for our galaxy m1e12 using data from all 5 snapshots taken at 100 Myr intervals in Fig. \ref{SurfaceDensPlot}. We see fluctuations in $X_{\mbox{\textsc{co}}}$ against $\Sigma_{\mbox{\textsc{h}}_{\oldstylenums{2}}}$, with a near constant $X_{\mbox{\textsc{co}}}$ between surface densities of 10$^{-6}$ to 100 M$_\odot$ pc$^{-2}$, but a sharp increase for surface densities > 100 M$_\odot$ pc$^{-2}$. This sharp increase aligns with the results found in \citet{shetty_modelling_2011-1}, as they also found an increase at surface densities > 100 M$_\odot$ pc$^{-2}$. This is because at high surface densities, the CO line is saturated, so the H$_2$ column density increases, as does the $X_{\mbox{\textsc{co}}}$ value. We can see that we see a similar upturn at 100 M$_\odot$ pc$^{-2}$ as seen in \citet{shetty_modelling_2011-1} which started at a similar gas surface density, which was attributed to the saturation of the CO line.

\begin{figure} 
    \includegraphics*[width=\columnwidth]{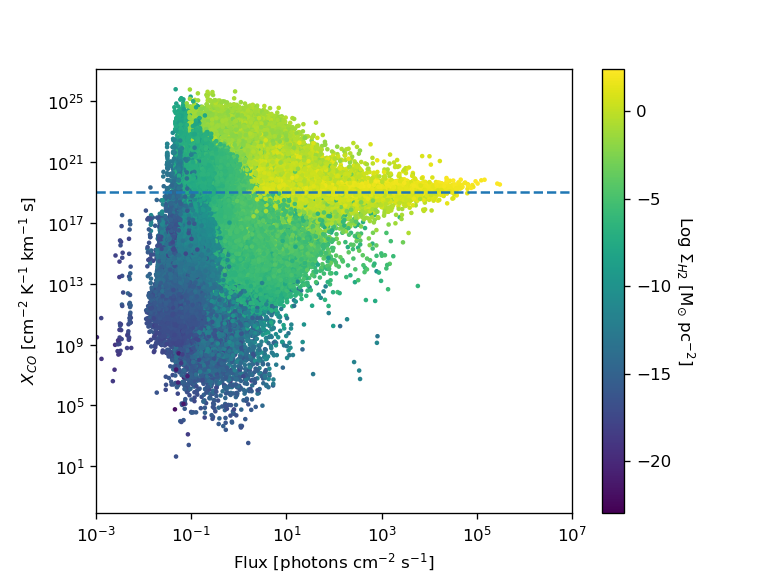}
    \caption{$X_{\mbox{\textsc{co}}}$ against FUV flux for each grid point with resolution of 20pc for galaxy m1e12 at time $t =$ 500 Myr, where we colour each point by $\Sigma_{\mbox{\textsc{h}}_{\oldstylenums{2}}}$. We see how areas with high-surface density and high-FUV flux correspond to a conversion factor similar to the value we find when averaging the $X_{\mbox{\textsc{co}}}$ of galaxy m1e12.} \label{XCOFluxPlot}
\end{figure}

We can also look at the strength of the far UV flux, which will correlate with star formation to see how $X_{\mbox{\textsc{co}}}$ changes when measuring the strength of the FUV flux. We show $X_{\mbox{\textsc{co}}}$ against FUV flux, and colour each point by $\Sigma_{\mbox{\textsc{h}}_{\oldstylenums{2}}}$ for galaxy m1e12 at time $t =$ 500 Myr in Figure \ref{XCOFluxPlot}. We see how $X_{\mbox{\textsc{co}}}$ changes with both $\Sigma_{\mbox{\textsc{h}}_{\oldstylenums{2}}}$ and the flux of the FUV field. We can also see that when the FUV flux and surface density are at their highest, the conversion factor matches what we measure on average in our simulations, showing how the conversion factor is dominated by areas of both high flux and molecular gas surface density, which corresponds to dense star forming molecular clouds within the galaxy. We also see how in regions with low FUV flux and low star formation, $X_{\mbox{\textsc{co}}}$ has a large scatter. However, as the FUV flux and star formation increase, this scatter is reduced and we see how the conversion factor is compacted around the average value we measure for the system.  





\section{Conclusion} \label{Conclusions}
We have analysed the CO emission from a suite of simulations based on the FIRE-2 subgrid models, and coupled them with the CHIMES chemistry module to evaluate the on-the-fly non-equilibrium chemistry against post-processing the abundances assuming chemical equilibrium. We then take snapshots from these galaxy models, and post-process through the \textsc{radmc} radiative transfer code, creating simulated spectra of the CO emission from our galaxy models. We find that:

\begin{enumerate}
    \item Our non-equilibrium fiducial models are able to match the relation seen in the Milky Way between the column densities of H$_2$ and CO. We show this in Figure \ref{gasCols}. We note that our Milky Way-mass galaxy m1e12 does not match the observations completely, as the median relation is $\approx$ 0.5 dex above the observations. However we believe this is due to the large metallicity increase that occurs within the initial 300 Myr of evolution of the model. We also further verify our H$_2$ model by comparing the fraction of molecular gas to the total neutral gas column density in Figure \ref{GasFracs}.
    
    \item Our simulations are able to match the CO luminosities as a function of star formation rate seen within the xCOLD GASS survey, which we show in Figure \ref{LCO_SFR}. We see our high-mass galaxies are \textasciitilde0.5 dex above the average for the xCOLD GASS survey, though still match to objects on the upper edge of the data for fixed SFR. For our low-metallicity galaxies outside the range of the xCOLD GASS survey, we note that they measure galaxies with SFR similar to our low-mass galaxy models and calculate a 3$\sigma$ upper limit on the CO luminosity for those. Future observations may be able to lower these upper limits and show whether our low-mass galaxies truly match.
    
    \item We plot our simulations against dust-based observations to evaluate the relation between metallicity and $X_{\mbox{\textsc{co}}}$ in Figure \ref{XCO_Z}. Our simulations follow the expected trend even at lower-metallicities, though we are unable to verify due to a lack of observational data due to the difficulties of measuring CO at low metallicity. 
    
    \item We evaluate the multivariate conversion factor from \citet{accurso_deriving_2017}, which in Figure \ref{MH2SFR} we see overpredicts the H$_2$ mass of our galaxy simulations by an order of magnitude. We follow the method used in \citet{accurso_deriving_2017} to find a new fitting to our simulations, which we show in Table \ref{FittingConstTable}. However, as we do not probe a full range of [C \textsc{ii}] luminosity and $\Delta$MS ranges, we see our simulations show a strange distribution when compared to a sample of the xCOLD GASS survey and DGS in Figure \ref{FittingObs} as some snapshots lay outside the observational data. For this reason, more simulations and/or observations are needed to reduce these caveats with our fitting function.
    
    \item We show how $X_{\mbox{\textsc{co}}}$ is mostly constant when applying a Gaussian filter across differing spatial scales to better match our simulated spectra to observations in Figure \ref{XCOGauss}. These slight variations we see in $X_{\mbox{\textsc{co}}}$ as the Gaussian filter FWHM increases are likely due to the Gaussian filter spreading H$_2$ gas or $W_{\mbox{\textsc{co}}}$ to regions where we see no gas or CO emission without the filter, which we can see in Figure \ref{GaussBlur_m1e12}.

    \item We compare $X_{\mbox{\textsc{co}}}$ to the surface density and the FUV flux, and show how $X_{\mbox{\textsc{co}}}$ within our simulations fluctuates with both in Figure \ref{XCOFluxPlot}. We also note that when the surface density and FUV flux is at its greatest, $X_{\mbox{\textsc{co}}}$ aligns with the conversion factor we measure in our m1e12 Milky Way-mass simulations (\textasciitilde 10$^{19}$ - 10$^{20}$ cm$^{-2}$(K km s$^{-1})^{-1}$).
    
    \item We quantify the effects non-equilibrium chemistry has on our galaxy simulations in table \ref{noneq_vs_eqm_L}. Whilst $L_{\mbox{\textsc{co}}}$ and $M_{\mbox{\textsc{h}}_{\oldstylenums{2}}}$ are reduced in equilibrium simulations by 24\% to 30\%, in higher-mass simulations these reduction factors are close, meaning $X_{\mbox{\textsc{co}}}$ in high-mass simulations is unaffected (<5\%) between equilibrium and non-equilibrium simulations
\end{enumerate}

Within this work, we have demonstrated that modelling non-equilibrium chemistry and synthetic emission line observations in hydrodynamical simulations is a promising avenue, allowing us to understand the CO-to-H$_2$ conversion factor in different environments. Whilst we see little difference in our $X_{\mbox{\textsc{co}}}$ estimates (<5\%) for our high-mass galaxies, we see how non-equilibrium chemistry can affect the simulated emission line as well as the molecular gas content of the models. However, this study is limited by our relatively small range of galaxy parameters spanned by our simulation suite. In future work, we will therefore expand the range of our simulations. One such way would be to look at $X_{\mbox{\textsc{co}}}$ in galaxy mergers, as we have only looked at isolated galaxies within this work, as this may reveal more about the nature of the conversion factor in extreme conditions. Collating the merger simulations, as well as our current simulations, could probe a greater parameter space than our current suite of simulations.


Another avenue for future work would be the use of machine learning to predict properties of galaxies and clouds. Recent work by \citet{garcia_textttslick_2023} designed \texttt{slick}, a package that calculates realistic CO luminosities for clouds and galaxies in hydrodynamical simulations using a model found through random forest machine learning methods. In future work, we will explore how machine learning techniques can be used to model the CO-to-H$_2$ conversion factor.

\section*{Acknowledgements}
We thank the reviewer for their report, and we thank Joop Schaye for useful comments and suggestions. CAFG was supported by NSF through grants AST-2108230, AST-2307327, and CAREER award AST-1652522; by NASA through grants 17-ATP17-0067 and 21-ATP21-0036; by STScI through grants HST-GO-16730.016-A and JWST-AR-03252.001-A; and by CXO through grant TM2-23005X. The simulations used in this work were run on the DiRAC@Durham facility managed by the Institute for Computational Cosmology on behalf of the STFC DiRAC HPC Facility (www.dirac.ac.uk). The equipment was funded by BEIS capital funding via STFC capital grants ST/K00042X/1, ST/P002293/1, ST/R002371/1, and ST/S002502/1, Durham University and STFC operations grant ST/R000832/1. DiRAC is part of the National eInfrastructure. The radiative transfer post-processing calculations and other analysis of the simulation outputs used Viper, the University of Hull High Performance Computing Facility.

\section*{Data Availability}
The data underlying this article will be shared on reasonable request to the corresponding author. A public version of the GIZMO code can be found at \href{http://www.tapir.caltech.edu/~phopkins/Site/GIZMO.html}{http://www.tapir.caltech.edu/\textasciitilde phopkins/Site/GIZMO.html}, and a public version of the CHIMES code can be found at \href{https://richings.bitbucket.io/chimes/home.html}{https://richings.bitbucket.io/chimes/home.html}.

\addcontentsline{toc}{section}{References}
\bibliography{main.bib}

\begin{thebibliography}{}
\makeatletter
\relax
\def\mn@urlcharsother{\let\do\@makeother \do\$\do\&\do\#\do\^\do\_\do\%\do\~}
\def\mn@doi{\begingroup\mn@urlcharsother \@ifnextchar [ {\mn@doi@} {\mn@doi@[]}}
\def\mn@doi@[#1]#2{\def\@tempa{#1}\ifx\@tempa\@empty \href {http://dx.doi.org/#2} {doi:#2}\else \href {http://dx.doi.org/#2} {#1}\fi \endgroup}
\def\mn@eprint#1#2{\mn@eprint@#1:#2::\@nil}
\def\mn@eprint@arXiv#1{\href {http://arxiv.org/abs/#1} {{\tt arXiv:#1}}}
\def\mn@eprint@dblp#1{\href {http://dblp.uni-trier.de/rec/bibtex/#1.xml} {dblp:#1}}
\def\mn@eprint@#1:#2:#3:#4\@nil{\def\@tempa {#1}\def\@tempb {#2}\def\@tempc {#3}\ifx \@tempc \@empty \let \@tempc \@tempb \let \@tempb \@tempa \fi \ifx \@tempb \@empty \def\@tempb {arXiv}\fi \@ifundefined {mn@eprint@\@tempb}{\@tempb:\@tempc}{\expandafter \expandafter \csname mn@eprint@\@tempb\endcsname \expandafter{\@tempc}}}

\bibitem[\protect\citeauthoryear{Accurso et~al.,}{Accurso et~al.}{2017}]{accurso_deriving_2017}
Accurso G.,  et~al., 2017, \mn@doi [Monthly Notices of the Royal Astronomical Society] {10.1093/mnras/stx1556}, 470, 4750

\bibitem[\protect\citeauthoryear{Allen, Hogg  \& Engelke}{Allen et~al.}{2015}]{allen_structure_2015}
Allen R.~J.,  Hogg D.~E.,   Engelke P.~D.,  2015, \mn@doi [The Astronomical Journal] {10.1088/0004-6256/149/4/123}, 149, 123

\bibitem[\protect\citeauthoryear{Amorín, Muñoz-Tuñón, Aguerri  \& Planesas}{Amorín et~al.}{2016}]{amorin_molecular_2016}
Amorín R.,  Muñoz-Tuñón C.,  Aguerri J. A.~L.,   Planesas P.,  2016, \mn@doi [Astronomy \& Astrophysics] {10.1051/0004-6361/201526397}, 588, A23

\bibitem[\protect\citeauthoryear{Andrews \& Martini}{Andrews \& Martini}{2013}]{andrews_mass-metallicity_2013}
Andrews B.~H.,  Martini P.,  2013, \mn@doi [The Astrophysical Journal] {10.1088/0004-637X/765/2/140}, 765, 140

\bibitem[\protect\citeauthoryear{Asano, Takeuchi, Hirashita  \& Inoue}{Asano et~al.}{2013}]{asano_dust_2013}
Asano R.~S.,  Takeuchi T.~T.,  Hirashita H.,   Inoue A.~K.,  2013, \mn@doi [Earth, Planets and Space] {10.5047/eps.2012.04.014}, 65, 213

\bibitem[\protect\citeauthoryear{Bekki}{Bekki}{2015}]{bekki_dust-regulated_2015}
Bekki K.,  2015, \mn@doi [Monthly Notices of the Royal Astronomical Society] {10.1093/mnras/stv165}, 449, 1625

\bibitem[\protect\citeauthoryear{Bolatto, Wolfire  \& Leroy}{Bolatto et~al.}{2013}]{bolatto_co--h_2013}
Bolatto A.~D.,  Wolfire M.,   Leroy A.~K.,  2013, \mn@doi [Annual Review of Astronomy and Astrophysics] {10.1146/annurev-astro-082812-140944}, 51, 207

\bibitem[\protect\citeauthoryear{Burgh, France  \& Jenkins}{Burgh et~al.}{2010}]{burgh_atomic_2010}
Burgh E.~B.,  France K.,   Jenkins E.~B.,  2010, \mn@doi [The Astrophysical Journal] {10.1088/0004-637X/708/1/334}, 708, 334

\bibitem[\protect\citeauthoryear{Castor}{Castor}{1970}]{castor_spectral_1970}
Castor J.~I.,  1970, \mn@doi [Monthly Notices of the Royal Astronomical Society] {10.1093/mnras/149.2.111}, 149, 111

\bibitem[\protect\citeauthoryear{Choban, Kereš, Hopkins, Sandstrom, Hayward  \& Faucher-Giguère}{Choban et~al.}{2022}]{choban_galactic_2022}
Choban C.~R.,  Kereš D.,  Hopkins P.~F.,  Sandstrom K.~M.,  Hayward C.~C.,   Faucher-Giguère C.-A.,  2022, \mn@doi [Monthly Notices of the Royal Astronomical Society] {10.1093/mnras/stac1542}, 514, 4506

\bibitem[\protect\citeauthoryear{Clark, Glover, Ragan  \& Duarte-Cabral}{Clark et~al.}{2019}]{clark_tracing_2019}
Clark P.~C.,  Glover S. C.~O.,  Ragan S.~E.,   Duarte-Cabral A.,  2019, \mn@doi [Monthly Notices of the Royal Astronomical Society] {10.1093/mnras/stz1119}, 486, 4622

\bibitem[\protect\citeauthoryear{Cormier et~al.,}{Cormier et~al.}{2014}]{cormier_molecular_2014}
Cormier D.,  et~al., 2014, \mn@doi [Astronomy \& Astrophysics] {10.1051/0004-6361/201322096}, 564, A121

\bibitem[\protect\citeauthoryear{Cormier et~al.,}{Cormier et~al.}{2015}]{cormier_herschel_2015}
Cormier D.,  et~al., 2015, \mn@doi [Astronomy \& Astrophysics] {10.1051/0004-6361/201425207}, 578, A53

\bibitem[\protect\citeauthoryear{Crenny \& Federman}{Crenny \& Federman}{2004}]{crenny_reanalysis_2004}
Crenny T.,  Federman S.~R.,  2004, \mn@doi [The Astrophysical Journal] {10.1086/382231}, 605, 278

\bibitem[\protect\citeauthoryear{D'Eugenio, Daddi, Liu  \& Gobat}{D'Eugenio et~al.}{2023}]{deugenio_cii_2023}
D'Eugenio C.,  Daddi E.,  Liu D.,   Gobat R.,  2023, \mn@doi [Astronomy \& Astrophysics] {10.1051/0004-6361/202347233}, 678, L9

\bibitem[\protect\citeauthoryear{De~Cia, Ledoux, Mattsson, Petitjean, Srianand, Gavignaud  \& Jenkins}{De~Cia et~al.}{2016}]{de_cia_dust-depletion_2016}
De~Cia A.,  Ledoux C.,  Mattsson L.,  Petitjean P.,  Srianand R.,  Gavignaud I.,   Jenkins E.~B.,  2016, \mn@doi [Astronomy \& Astrophysics] {10.1051/0004-6361/201527895}, 596, A97

\bibitem[\protect\citeauthoryear{Dere, Landi, Mason, Monsignori~Fossi  \& Young}{Dere et~al.}{1997}]{dere_chianti_1997}
Dere K.~P.,  Landi E.,  Mason H.~E.,  Monsignori~Fossi B.~C.,   Young P.~R.,  1997, \mn@doi [Astronomy and Astrophysics Supplement Series] {10.1051/aas:1997368}, 125, 149

\bibitem[\protect\citeauthoryear{Dullemond, Juhasz, Pohl, Sereshti, Shetty, Peters, Commercon  \& Flock}{Dullemond et~al.}{2012}]{dullemond_radmc-3d_2012}
Dullemond C.~P.,  Juhasz A.,  Pohl A.,  Sereshti F.,  Shetty R.,  Peters T.,  Commercon B.,   Flock M.,  2012, Astrophysics Source Code Library, record ascl:1202.015

\bibitem[\protect\citeauthoryear{Federman, Huntress  \& Prasad}{Federman et~al.}{1990}]{federman_modeling_1990}
Federman S.~R.,  Huntress Jr. W.~T.,   Prasad S.~S.,  1990, \mn@doi [The Astrophysical Journal] {10.1086/168711}, 354, 504

\bibitem[\protect\citeauthoryear{Feldmann, Gnedin  \& Kravtsov}{Feldmann et~al.}{2011}]{feldmann_how_2011}
Feldmann R.,  Gnedin N.~Y.,   Kravtsov A.~V.,  2011, \mn@doi [The Astrophysical Journal] {10.1088/0004-637X/732/2/115}, 732, 115

\bibitem[\protect\citeauthoryear{Feldmann, Gnedin  \& Kravtsov}{Feldmann et~al.}{2012a}]{feldmann_x-factor_2012-1}
Feldmann R.,  Gnedin N.~Y.,   Kravtsov A.~V.,  2012a, \mn@doi [The Astrophysical Journal] {10.1088/0004-637X/747/2/124}, 747, 124

\bibitem[\protect\citeauthoryear{Feldmann, Gnedin  \& Kravtsov}{Feldmann et~al.}{2012b}]{feldmann_x-factor_2012}
Feldmann R.,  Gnedin N.~Y.,   Kravtsov A.~V.,  2012b, \mn@doi [The Astrophysical Journal] {10.1088/0004-637X/758/2/127}, 758, 127

\bibitem[\protect\citeauthoryear{Ferland et~al.,}{Ferland et~al.}{2013}]{ferland_2013_2013}
Ferland G.~J.,  et~al., 2013, \mn@doi [RMxAA] {10.48550/ARXIV.1302.4485}, 49, 137

\bibitem[\protect\citeauthoryear{Flores Velázquez et~al.,}{Flores Velázquez et~al.}{2021}]{floresvelazquez_time-scales_2021}
Flores Velázquez J.~A.,  et~al., 2021, \mn@doi [Monthly Notices of the Royal Astronomical Society] {10.1093/mnras/staa3893}, 501, 4812

\bibitem[\protect\citeauthoryear{Fujimoto et~al.,}{Fujimoto et~al.}{2024}]{fujimoto_jwst_2024}
Fujimoto S.,  et~al., 2024, \mn@doi [The Astrophysical Journal] {10.3847/1538-4357/ad235c}, 964, 146

\bibitem[\protect\citeauthoryear{Garcia, Narayanan, Popping, Anirudh, Sutherland  \& Kaasinen}{Garcia et~al.}{2023}]{garcia_textttslick_2023}
Garcia K.,  Narayanan D.,  Popping G.,  Anirudh R.,  Sutherland S.,   Kaasinen M.,  2023, \${\textbackslash}texttt\{slick\}\$: {Modeling} a {Universe} of {Molecular} {Line} {Luminosities} in {Hydrodynamical} {Simulations}, \url {http://arxiv.org/abs/2311.01508}

\bibitem[\protect\citeauthoryear{Genzel et~al.,}{Genzel et~al.}{2012}]{genzel_metallicity_2012}
Genzel R.,  et~al., 2012, \mn@doi [The Astrophysical Journal] {10.1088/0004-637X/746/1/69}, 746, 69

\bibitem[\protect\citeauthoryear{Gerin \& Phillips}{Gerin \& Phillips}{2000}]{gerin_atomic_2000}
Gerin M.,  Phillips T.~G.,  2000, \mn@doi [The Astrophysical Journal] {10.1086/309072}, 537, 644

\bibitem[\protect\citeauthoryear{Gillmon, Shull, Tumlinson  \& Danforth}{Gillmon et~al.}{2006}]{gillmon_fuse_2006}
Gillmon K.,  Shull J.~M.,  Tumlinson J.,   Danforth C.,  2006, \mn@doi [The Astrophysical Journal] {10.1086/498053}, 636, 891

\bibitem[\protect\citeauthoryear{Glover \& Clark}{Glover \& Clark}{2016}]{glover_is_2016}
Glover S. C.~O.,  Clark P.~C.,  2016, \mn@doi [Monthly Notices of the Royal Astronomical Society] {10.1093/mnras/stv2863}, 456, 3596

\bibitem[\protect\citeauthoryear{Glover, Clark, Micic  \& Molina}{Glover et~al.}{2015}]{glover_modelling_2015}
Glover S. C.~O.,  Clark P.~C.,  Micic M.,   Molina F.,  2015, \mn@doi [Monthly Notices of the Royal Astronomical Society] {10.1093/mnras/stu2699}, 448, 1607

\bibitem[\protect\citeauthoryear{Gnedin \& Kravtsov}{Gnedin \& Kravtsov}{2011}]{gnedin_environmental_2011}
Gnedin N.~Y.,  Kravtsov A.~V.,  2011, \mn@doi [The Astrophysical Journal] {10.1088/0004-637X/728/2/88}, 728, 88

\bibitem[\protect\citeauthoryear{Gnedin, Tassis  \& Kravtsov}{Gnedin et~al.}{2009}]{gnedin_modeling_2009}
Gnedin N.~Y.,  Tassis K.,   Kravtsov A.~V.,  2009, \mn@doi [The Astrophysical Journal] {10.1088/0004-637X/697/1/55}, 697, 55

\bibitem[\protect\citeauthoryear{Goldreich \& Kwan}{Goldreich \& Kwan}{1974}]{goldreich_molecular_1974}
Goldreich P.,  Kwan J.,  1974, \mn@doi [The Astrophysical Journal] {10.1086/152821}, 189, 441

\bibitem[\protect\citeauthoryear{Gong, Ostriker, Kim  \& Kim}{Gong et~al.}{2020}]{gong_environmental_2020}
Gong M.,  Ostriker E.~C.,  Kim C.-G.,   Kim J.-G.,  2020, \mn@doi [The Astrophysical Journal] {10.3847/1538-4357/abbdab}, 903, 142

\bibitem[\protect\citeauthoryear{Grenier, Casandjian  \& Terrier}{Grenier et~al.}{2005}]{grenier_unveiling_2005}
Grenier I.~A.,  Casandjian J.-M.,   Terrier R.,  2005, \mn@doi [Science] {10.1126/science.1106924}, 307, 1292

\bibitem[\protect\citeauthoryear{Gurvich}{Gurvich}{2022}]{gurvich_fire_2022}
Gurvich A.~B.,  2022, Astrophysics Source Code Library, record ascl:2202.006

\bibitem[\protect\citeauthoryear{Hirashita}{Hirashita}{2023}]{hirashita_effects_2023}
Hirashita H.,  2023, \mn@doi [Monthly Notices of the Royal Astronomical Society] {10.1093/mnras/stad1286}, 522, 4612

\bibitem[\protect\citeauthoryear{Hirashita, Nozawa, Villaume  \& Srinivasan}{Hirashita et~al.}{2015}]{hirashita_dust_2015}
Hirashita H.,  Nozawa T.,  Villaume A.,   Srinivasan S.,  2015, \mn@doi [Monthly Notices of the Royal Astronomical Society] {10.1093/mnras/stv2095}, 454, 1620

\bibitem[\protect\citeauthoryear{Hopkins}{Hopkins}{2015}]{hopkins_new_2015}
Hopkins P.~F.,  2015, \mn@doi [Monthly Notices of the Royal Astronomical Society] {10.1093/mnras/stv195}, 450, 53

\bibitem[\protect\citeauthoryear{Hopkins et~al.,}{Hopkins et~al.}{2018a}]{hopkins_how_2018}
Hopkins P.~F.,  et~al., 2018a, \mn@doi [Monthly Notices of the Royal Astronomical Society] {10.1093/mnras/sty674}, 477, 1578

\bibitem[\protect\citeauthoryear{Hopkins et~al.,}{Hopkins et~al.}{2018b}]{hopkins_fire-2_2018}
Hopkins P.~F.,  et~al., 2018b, \mn@doi [Monthly Notices of the Royal Astronomical Society] {10.1093/mnras/sty1690}, 480, 800

\bibitem[\protect\citeauthoryear{Hopkins, Grudić, Wetzel, Kereš, Faucher-Giguère, Ma, Murray  \& Butcher}{Hopkins et~al.}{2020}]{hopkins_radiative_2020}
Hopkins P.~F.,  Grudić M.~Y.,  Wetzel A.,  Kereš D.,  Faucher-Giguère C.-A.,  Ma X.,  Murray N.,   Butcher N.,  2020, \mn@doi [Monthly Notices of the Royal Astronomical Society] {10.1093/mnras/stz3129}, 491, 3702

\bibitem[\protect\citeauthoryear{Hu, Schruba, Sternberg  \& Van~Dishoeck}{Hu et~al.}{2022}]{hu__dependence_2022}
Hu C.-Y.,  Schruba A.,  Sternberg A.,   Van~Dishoeck E.~F.,  2022, \mn@doi [The Astrophysical Journal] {10.3847/1538-4357/ac65fd}, 931, 28

\bibitem[\protect\citeauthoryear{{Hu}, {Sternberg}  \& {van Dishoeck}}{{Hu} et~al.}{2023}]{hu_co-evolution_2023}
{Hu} C.-Y.,  {Sternberg} A.,   {van Dishoeck} E.~F.,  2023, \mn@doi [\apj] {10.3847/1538-4357/acdcfa}, \href {https://ui.adsabs.harvard.edu/abs/2023ApJ...952..140H} {952, 140}

\bibitem[\protect\citeauthoryear{Israel}{Israel}{1997}]{israel_h2_1997}
Israel F.~P.,  1997, \mn@doi [Astronomy and Astrophysics] {10.48550/ARXIV.ASTRO-PH/9709194}, 328, 471

\bibitem[\protect\citeauthoryear{Iwamoto, Brachwitz, Nomoto, Kishimoto, Umeda, Hix  \& Thielemann}{Iwamoto et~al.}{1999}]{iwamoto_nucleosynthesis_1999}
Iwamoto K.,  Brachwitz F.,  Nomoto K.,  Kishimoto N.,  Umeda H.,  Hix W.~R.,   Thielemann F.,  1999, \mn@doi [The Astrophysical Journal Supplement Series] {10.1086/313278}, 125, 439

\bibitem[\protect\citeauthoryear{Izzard, Tout, Karakas  \& Pols}{Izzard et~al.}{2004}]{izzard_new_2004}
Izzard R.~G.,  Tout C.~A.,  Karakas A.~I.,   Pols O.~R.,  2004, \mn@doi [Monthly Notices of the Royal Astronomical Society] {10.1111/j.1365-2966.2004.07446.x}, 350, 407

\bibitem[\protect\citeauthoryear{Janowiecki, Catinella, Cortese, Saintonge, Brown  \& Wang}{Janowiecki et~al.}{2017}]{janowiecki_xgass_2017}
Janowiecki S.,  Catinella B.,  Cortese L.,  Saintonge A.,  Brown T.,   Wang J.,  2017, \mn@doi [Monthly Notices of the Royal Astronomical Society] {10.1093/mnras/stx046}, p. stx046

\bibitem[\protect\citeauthoryear{Jenkins}{Jenkins}{2009}]{jenkins_unified_2009}
Jenkins E.~B.,  2009, \mn@doi [The Astrophysical Journal] {10.1088/0004-637X/700/2/1299}, 700, 1299

\bibitem[\protect\citeauthoryear{Keating et~al.,}{Keating et~al.}{2020}]{keating_reproducing_2020}
Keating L.~C.,  et~al., 2020, \mn@doi [Monthly Notices of the Royal Astronomical Society] {10.1093/mnras/staa2839}, 499, 837

\bibitem[\protect\citeauthoryear{Kennicutt \& Evans}{Kennicutt \& Evans}{2012}]{kennicutt_star_2012}
Kennicutt R.~C.,  Evans N.~J.,  2012, \mn@doi [Annual Review of Astronomy and Astrophysics] {10.1146/annurev-astro-081811-125610}, 50, 531

\bibitem[\protect\citeauthoryear{Kim \& Ostriker}{Kim \& Ostriker}{2017}]{kim_three-phase_2017}
Kim C.-G.,  Ostriker E.~C.,  2017, \mn@doi [The Astrophysical Journal] {10.3847/1538-4357/aa8599}, 846, 133

\bibitem[\protect\citeauthoryear{Kroupa}{Kroupa}{2001}]{kroupa_variation_2001}
Kroupa P.,  2001, \mn@doi [Monthly Notices of the Royal Astronomical Society] {10.1046/j.1365-8711.2001.04022.x}, 322, 231

\bibitem[\protect\citeauthoryear{Landi, Young, Dere, Del~Zanna  \& Mason}{Landi et~al.}{2013}]{landi_chiantiatomic_2013}
Landi E.,  Young P.~R.,  Dere K.~P.,  Del~Zanna G.,   Mason H.~E.,  2013, \mn@doi [The Astrophysical Journal] {10.1088/0004-637X/763/2/86}, 763, 86

\bibitem[\protect\citeauthoryear{Leitherer et~al.,}{Leitherer et~al.}{1999}]{leitherer_starburst99_1999}
Leitherer C.,  et~al., 1999, \mn@doi [The Astrophysical Journal Supplement Series] {10.1086/313233}, 123, 3

\bibitem[\protect\citeauthoryear{Leroy, Bolatto, Simon  \& Blitz}{Leroy et~al.}{2005}]{leroy_molecular_2005}
Leroy A.,  Bolatto A.~D.,  Simon J.~D.,   Blitz L.,  2005, \mn@doi [The Astrophysical Journal] {10.1086/429578}, 625, 763

\bibitem[\protect\citeauthoryear{Leroy et~al.,}{Leroy et~al.}{2011}]{leroy_co--h_2011}
Leroy A.~K.,  et~al., 2011, \mn@doi [The Astrophysical Journal] {10.1088/0004-637X/737/1/12}, 737, 12

\bibitem[\protect\citeauthoryear{Leroy et~al.,}{Leroy et~al.}{2021}]{leroy_phangsalma_2021}
Leroy A.~K.,  et~al., 2021, \mn@doi [The Astrophysical Journal Supplement Series] {10.3847/1538-4365/ac17f3}, 257, 43

\bibitem[\protect\citeauthoryear{Levrier, Le~Petit, Hennebelle, Lesaffre, Gerin  \& Falgarone}{Levrier et~al.}{2012}]{levrier_uv-driven_2012}
Levrier F.,  Le~Petit F.,  Hennebelle P.,  Lesaffre P.,  Gerin M.,   Falgarone E.,  2012, \mn@doi [Astronomy \& Astrophysics] {10.1051/0004-6361/201218865}, 544, A22

\bibitem[\protect\citeauthoryear{Li, Narayanan, Davè  \& Krumholz}{Li et~al.}{2018}]{li_dark_2018}
Li Q.,  Narayanan D.,  Davè R.,   Krumholz M.~R.,  2018, \mn@doi [The Astrophysical Journal] {10.3847/1538-4357/aaec77}, 869, 73

\bibitem[\protect\citeauthoryear{Liszt \& Lucas}{Liszt \& Lucas}{1996}]{liszt_galactic_1996}
Liszt H.,  Lucas R.,  1996, Astronomy \& Astrophysics, 314

\bibitem[\protect\citeauthoryear{Madden, Poglitsch, Geis, Stacey  \& Townes}{Madden et~al.}{1997}]{madden_c_1997}
Madden S.~C.,  Poglitsch A.,  Geis N.,  Stacey G.~J.,   Townes C.~H.,  1997, \mn@doi [The Astrophysical Journal] {10.1086/304247}, 483, 200

\bibitem[\protect\citeauthoryear{Madden et~al.,}{Madden et~al.}{2013}]{madden_overview_2013}
Madden S.~C.,  et~al., 2013, \mn@doi [Publications of the Astronomical Society of the Pacific] {10.1086/671138}, 125, 600

\bibitem[\protect\citeauthoryear{Madden et~al.,}{Madden et~al.}{2020}]{madden_tracing_2020}
Madden S.~C.,  et~al., 2020, \mn@doi [Astronomy \& Astrophysics] {10.1051/0004-6361/202038860}, 643, A141

\bibitem[\protect\citeauthoryear{Mannucci, Della~Valle  \& Panagia}{Mannucci et~al.}{2006}]{mannucci_two_2006}
Mannucci F.,  Della~Valle M.,   Panagia N.,  2006, \mn@doi [Monthly Notices of the Royal Astronomical Society] {10.1111/j.1365-2966.2006.10501.x}, 370, 773

\bibitem[\protect\citeauthoryear{Marigo}{Marigo}{2001}]{marigo_chemical_2001}
Marigo P.,  2001, \mn@doi [Astronomy \& Astrophysics] {10.1051/0004-6361:20000247}, 370, 194

\bibitem[\protect\citeauthoryear{Mathis, Rumpl  \& Nordsieck}{Mathis et~al.}{1977}]{mathis_size_1977}
Mathis J.~S.,  Rumpl W.,   Nordsieck K.~H.,  1977, \mn@doi [The Astrophysical Journal] {10.1086/155591}, 217, 425

\bibitem[\protect\citeauthoryear{McKinnon, Vogelsberger, Torrey, Marinacci  \& Kannan}{McKinnon et~al.}{2018}]{mckinnon_simulating_2018}
McKinnon R.,  Vogelsberger M.,  Torrey P.,  Marinacci F.,   Kannan R.,  2018, \mn@doi [Monthly Notices of the Royal Astronomical Society] {10.1093/mnras/sty1248}, 478, 2851

\bibitem[\protect\citeauthoryear{Moster, Naab  \& White}{Moster et~al.}{2013}]{moster_galactic_2013}
Moster B.~P.,  Naab T.,   White S. D.~M.,  2013, \mn@doi [Monthly Notices of the Royal Astronomical Society] {10.1093/mnras/sts261}, 428, 3121

\bibitem[\protect\citeauthoryear{Narayanan, Krumholz, Ostriker  \& Hernquist}{Narayanan et~al.}{2011}]{narayanan_co-h2_2011}
Narayanan D.,  Krumholz M.,  Ostriker E.~C.,   Hernquist L.,  2011, \mn@doi [Monthly Notices of the Royal Astronomical Society] {10.1111/j.1365-2966.2011.19516.x}, 418, 664

\bibitem[\protect\citeauthoryear{Narayanan, Krumholz, Ostriker  \& Hernquist}{Narayanan et~al.}{2012}]{narayanan_general_2012}
Narayanan D.,  Krumholz M.~R.,  Ostriker E.~C.,   Hernquist L.,  2012, \mn@doi [Monthly Notices of the Royal Astronomical Society] {10.1111/j.1365-2966.2012.20536.x}, 421, 3127

\bibitem[\protect\citeauthoryear{Nguyen et~al.,}{Nguyen et~al.}{2018}]{nguyen_dustgas_2018}
Nguyen H.,  et~al., 2018, \mn@doi [The Astrophysical Journal] {10.3847/1538-4357/aac82b}, 862, 49

\bibitem[\protect\citeauthoryear{Nomoto, Tominaga, Umeda, Kobayashi  \& Maeda}{Nomoto et~al.}{2006}]{nomoto_nucleosynthesis_2006}
Nomoto K.,  Tominaga N.,  Umeda H.,  Kobayashi C.,   Maeda K.,  2006, \mn@doi [Nuclear Physics A] {10.1016/j.nuclphysa.2006.05.008}, 777, 424

\bibitem[\protect\citeauthoryear{Offner, Bisbas, Bell  \& Viti}{Offner et~al.}{2014}]{offner_alternative_2014}
Offner S. S.~R.,  Bisbas T.~G.,  Bell T.~A.,   Viti S.,  2014, \mn@doi [Monthly Notices of the Royal Astronomical Society: Letters] {10.1093/mnrasl/slu013}, 440, L81

\bibitem[\protect\citeauthoryear{Orr et~al.,}{Orr et~al.}{2018}]{orr_what_2018}
Orr M.~E.,  et~al., 2018, \mn@doi [Monthly Notices of the Royal Astronomical Society] {10.1093/mnras/sty1241}, 478, 3653

\bibitem[\protect\citeauthoryear{Papadopoulos, Thi  \& Viti}{Papadopoulos et~al.}{2004}]{papadopoulos_c_2004}
Papadopoulos P.~P.,  Thi W.-F.,   Viti S.,  2004, \mn@doi [Monthly Notices of the Royal Astronomical Society] {10.1111/j.1365-2966.2004.07762.x}, 351, 147

\bibitem[\protect\citeauthoryear{Pilbratt et~al.,}{Pilbratt et~al.}{2010}]{pilbratt_herschel_2010}
Pilbratt G.~L.,  et~al., 2010, \mn@doi [Astronomy and Astrophysics] {10.1051/0004-6361/201014759}, 518, L1

\bibitem[\protect\citeauthoryear{Poglitsch, Krabbe, Madden, Nikola, Geis, Johansson, Stacey  \& Sternberg}{Poglitsch et~al.}{1995}]{poglitsch_multiwavelength_1995}
Poglitsch A.,  Krabbe A.,  Madden S.~C.,  Nikola T.,  Geis N.,  Johansson L. E.~B.,  Stacey G.~J.,   Sternberg A.,  1995, \mn@doi [The Astrophysical Journal] {10.1086/176482}, 454, 293

\bibitem[\protect\citeauthoryear{Poglitsch et~al.,}{Poglitsch et~al.}{2010}]{poglitsch_photodetector_2010}
Poglitsch A.,  et~al., 2010, \mn@doi [Astronomy and Astrophysics] {10.1051/0004-6361/201014535}, 518, L2

\bibitem[\protect\citeauthoryear{Rachford et~al.,}{Rachford et~al.}{2002}]{rachford_far_2002}
Rachford B.~L.,  et~al., 2002, \mn@doi [The Astrophysical Journal] {10.1086/342146}, 577, 221

\bibitem[\protect\citeauthoryear{Ramambason et~al.,}{Ramambason et~al.}{2024}]{ramambason_modeling_2024}
Ramambason L.,  et~al., 2024, \mn@doi [Astronomy \& Astrophysics] {10.1051/0004-6361/202347280}, 681, A14

\bibitem[\protect\citeauthoryear{Richings \& Schaye}{Richings \& Schaye}{2016a}]{richings_effects_2016}
Richings A.~J.,  Schaye J.,  2016a, \mn@doi [Monthly Notices of the Royal Astronomical Society] {10.1093/mnras/stw327}, 458, 270

\bibitem[\protect\citeauthoryear{Richings \& Schaye}{Richings \& Schaye}{2016b}]{richings_chemical_2016}
Richings A.~J.,  Schaye J.,  2016b, \mn@doi [Monthly Notices of the Royal Astronomical Society] {10.1093/mnras/stw1135}, 460, 2297

\bibitem[\protect\citeauthoryear{Richings, Schaye  \& Oppenheimer}{Richings et~al.}{2014a}]{richings_non-equilibrium_2014-1}
Richings A.~J.,  Schaye J.,   Oppenheimer B.~D.,  2014a, \mn@doi [Monthly Notices of the Royal Astronomical Society] {10.1093/mnras/stu525}, 440, 3349

\bibitem[\protect\citeauthoryear{Richings, Schaye  \& Oppenheimer}{Richings et~al.}{2014b}]{richings_non-equilibrium_2014}
Richings A.~J.,  Schaye J.,   Oppenheimer B.~D.,  2014b, \mn@doi [Monthly Notices of the Royal Astronomical Society] {10.1093/mnras/stu1046}, 442, 2780

\bibitem[\protect\citeauthoryear{Richings, Faucher-Giguère, Gurvich, Schaye  \& Hayward}{Richings et~al.}{2022}]{richings_effects_2022}
Richings A.~J.,  Faucher-Giguère C.-A.,  Gurvich A.~B.,  Schaye J.,   Hayward C.~C.,  2022, \mn@doi [Monthly Notices of the Royal Astronomical Society] {10.1093/mnras/stac2338}, 517, 1557

\bibitem[\protect\citeauthoryear{Rubio, Elmegreen, Hunter, Brinks, Cortés  \& Cigan}{Rubio et~al.}{2015}]{rubio_dense_2015}
Rubio M.,  Elmegreen B.~G.,  Hunter D.~A.,  Brinks E.,  Cortés J.~R.,   Cigan P.,  2015, \mn@doi [Nature] {10.1038/nature14901}, 525, 218

\bibitem[\protect\citeauthoryear{Rémy-Ruyer et~al.,}{Rémy-Ruyer et~al.}{2014}]{remy-ruyer_gas--dust_2014}
Rémy-Ruyer A.,  et~al., 2014, \mn@doi [Astronomy \& Astrophysics] {10.1051/0004-6361/201322803}, 563, A31

\bibitem[\protect\citeauthoryear{Saintonge et~al.,}{Saintonge et~al.}{2017}]{saintonge_xcold_2017}
Saintonge A.,  et~al., 2017, \mn@doi [The Astrophysical Journal Supplement Series] {10.3847/1538-4365/aa97e0}, 233, 22

\bibitem[\protect\citeauthoryear{Sandstrom et~al.,}{Sandstrom et~al.}{2013}]{sandstrom_co--h_2013}
Sandstrom K.~M.,  et~al., 2013, \mn@doi [The Astrophysical Journal] {10.1088/0004-637X/777/1/5}, 777, 5

\bibitem[\protect\citeauthoryear{Sawala et~al.,}{Sawala et~al.}{2015}]{sawala_bent_2015}
Sawala T.,  et~al., 2015, \mn@doi [Monthly Notices of the Royal Astronomical Society] {10.1093/mnras/stu2753}, 448, 2941

\bibitem[\protect\citeauthoryear{Schilke, Phillips  \& Wang}{Schilke et~al.}{1995}]{schilke_hydrogen_1995}
Schilke P.,  Phillips T.~G.,   Wang N.,  1995, \mn@doi [The Astrophysical Journal] {10.1086/175358}, 441, 334

\bibitem[\protect\citeauthoryear{Schruba et~al.,}{Schruba et~al.}{2012}]{schruba_low_2012}
Schruba A.,  et~al., 2012, \mn@doi [The Astronomical Journal] {10.1088/0004-6256/143/6/138}, 143, 138

\bibitem[\protect\citeauthoryear{Schruba et~al.,}{Schruba et~al.}{2017}]{schruba_physical_2017}
Schruba A.,  et~al., 2017, \mn@doi [The Astrophysical Journal] {10.3847/1538-4357/835/2/278}, 835, 278

\bibitem[\protect\citeauthoryear{Schöier, Van Der~Tak, Van~Dishoeck  \& Black}{Schöier et~al.}{2005}]{schoier_atomic_2005}
Schöier F.~L.,  Van Der~Tak F. F.~S.,  Van~Dishoeck E.~F.,   Black J.~H.,  2005, \mn@doi [Astronomy \& Astrophysics] {10.1051/0004-6361:20041729}, 432, 369

\bibitem[\protect\citeauthoryear{Seifried et~al.,}{Seifried et~al.}{2017}]{seifried_silcc-zoom_2017}
Seifried D.,  et~al., 2017, \mn@doi [Monthly Notices of the Royal Astronomical Society] {10.1093/mnras/stx2343}, 472, 4797

\bibitem[\protect\citeauthoryear{Seifried, Haid, Walch, Borchert  \& Bisbas}{Seifried et~al.}{2020}]{seifried_silcc-zoom_2020}
Seifried D.,  Haid S.,  Walch S.,  Borchert E. M.~A.,   Bisbas T.~G.,  2020, \mn@doi [Monthly Notices of the Royal Astronomical Society] {10.1093/mnras/stz3563}, 492, 1465

\bibitem[\protect\citeauthoryear{Sheffer, Rogers, Federman, Abel, Gredel, Lambert  \& Shaw}{Sheffer et~al.}{2008}]{sheffer_ultraviolet_2008}
Sheffer Y.,  Rogers M.,  Federman S.~R.,  Abel N.~P.,  Gredel R.,  Lambert D.~L.,   Shaw G.,  2008, \mn@doi [The Astrophysical Journal] {10.1086/591484}, 687, 1075

\bibitem[\protect\citeauthoryear{Shetty, Glover, Dullemond  \& Klessen}{Shetty et~al.}{2011a}]{shetty_modelling_2011}
Shetty R.,  Glover S.~C.,  Dullemond C.~P.,   Klessen R.~S.,  2011a, \mn@doi [Monthly Notices of the Royal Astronomical Society] {10.1111/j.1365-2966.2010.18005.x}, 412, 1686

\bibitem[\protect\citeauthoryear{Shetty, Glover, Dullemond, Ostriker, Harris  \& Klessen}{Shetty et~al.}{2011b}]{shetty_modelling_2011-1}
Shetty R.,  Glover S.~C.,  Dullemond C.~P.,  Ostriker E.~C.,  Harris A.~I.,   Klessen R.~S.,  2011b, \mn@doi [Monthly Notices of the Royal Astronomical Society] {10.1111/j.1365-2966.2011.18937.x}, 415, 3253

\bibitem[\protect\citeauthoryear{Shull, Danforth  \& Anderson}{Shull et~al.}{2021}]{shull_far_2021}
Shull J.~M.,  Danforth C.~W.,   Anderson K.~L.,  2021, \mn@doi [The Astrophysical Journal] {10.3847/1538-4357/abe707}, 911, 55

\bibitem[\protect\citeauthoryear{Sofia, Parvathi, Babu  \& Murthy}{Sofia et~al.}{2011}]{sofia_determining_2011}
Sofia U.~J.,  Parvathi V.~S.,  Babu B. R.~S.,   Murthy J.,  2011, \mn@doi [The Astronomical Journal] {10.1088/0004-6256/141/1/22}, 141, 22

\bibitem[\protect\citeauthoryear{Solomon, Downes, Radford  \& Barrett}{Solomon et~al.}{1997}]{solomon_molecular_1997}
Solomon P.~M.,  Downes D.,  Radford S. J.~E.,   Barrett J.~W.,  1997, \mn@doi [The Astrophysical Journal] {10.1086/303765}, 478, 144

\bibitem[\protect\citeauthoryear{Teng et~al.,}{Teng et~al.}{2024}]{teng_star_2024}
Teng Y.-H.,  et~al., 2024, \mn@doi [The Astrophysical Journal] {10.3847/1538-4357/ad10ae}, 961, 42

\bibitem[\protect\citeauthoryear{Tsai \& Mathews}{Tsai \& Mathews}{1995}]{tsai_interstellar_1995}
Tsai J.~C.,  Mathews W.~G.,  1995, \mn@doi [The Astrophysical Journal] {10.1086/175943}, 448, 84

\bibitem[\protect\citeauthoryear{Tumlinson et~al.,}{Tumlinson et~al.}{2002}]{tumlinson_far_2002}
Tumlinson J.,  et~al., 2002, \mn@doi [The Astrophysical Journal] {10.1086/338112}, 566, 857

\bibitem[\protect\citeauthoryear{Van Den~Hoek \& Groenewegen}{Van Den~Hoek \& Groenewegen}{1997}]{van_den_hoek_new_1997}
Van Den~Hoek L.~B.,  Groenewegen M.~A.,  1997, \mn@doi [Astronomy and Astrophysics Supplement Series] {10.1051/aas:1997162}, 123, 305

\bibitem[\protect\citeauthoryear{Van~Dishoeck \& Black}{Van~Dishoeck \& Black}{1988}]{van_dishoeck_photodissociation_1988}
Van~Dishoeck E.~F.,  Black J.~H.,  1988, \mn@doi [The Astrophysical Journal] {10.1086/166877}, 334, 771

\bibitem[\protect\citeauthoryear{Vizgan et~al.,}{Vizgan et~al.}{2022}]{vizgan_tracing_2022}
Vizgan D.,  et~al., 2022, \mn@doi [The Astrophysical Journal] {10.3847/1538-4357/ac5cba}, 929, 92

\bibitem[\protect\citeauthoryear{Walch et~al.,}{Walch et~al.}{2015}]{walch_silcc_2015}
Walch S.,  et~al., 2015, \mn@doi [Monthly Notices of the Royal Astronomical Society] {10.1093/mnras/stv1975}, 454, 246

\bibitem[\protect\citeauthoryear{Wiersma, Schaye  \& Smith}{Wiersma et~al.}{2009}]{wiersma_effect_2009}
Wiersma R. P.~C.,  Schaye J.,   Smith B.~D.,  2009, \mn@doi [Monthly Notices of the Royal Astronomical Society] {10.1111/j.1365-2966.2008.14191.x}, 393, 99

\bibitem[\protect\citeauthoryear{Wilson}{Wilson}{1995}]{wilson_metallicity_1995}
Wilson C.~D.,  1995, \mn@doi [The Astrophysical Journal] {10.1086/309615}, 448

\bibitem[\protect\citeauthoryear{Wolfire, Tielens, Hollenbach  \& Kaufman}{Wolfire et~al.}{2008}]{wolfire_chemical_2008}
Wolfire M.~G.,  Tielens A. G. G.~M.,  Hollenbach D.,   Kaufman M.~J.,  2008, \mn@doi [The Astrophysical Journal] {10.1086/587688}, 680, 384

\bibitem[\protect\citeauthoryear{Wolfire, Hollenbach  \& McKee}{Wolfire et~al.}{2010}]{wolfire_dark_2010}
Wolfire M.~G.,  Hollenbach D.,   McKee C.~F.,  2010, \mn@doi [The Astrophysical Journal] {10.1088/0004-637X/716/2/1191}, 716, 1191

\bibitem[\protect\citeauthoryear{Xu, Li, Yue  \& Goldsmith}{Xu et~al.}{2016}]{xu_evolution_2016}
Xu D.,  Li D.,  Yue N.,   Goldsmith P.~F.,  2016, \mn@doi [The Astrophysical Journal] {10.3847/0004-637X/819/1/22}, 819, 22

\bibitem[\protect\citeauthoryear{Zanella et~al.,}{Zanella et~al.}{2018}]{zanella_c_2018}
Zanella A.,  et~al., 2018, \mn@doi [Monthly Notices of the Royal Astronomical Society] {10.1093/mnras/sty2394}, 481, 1976

\bibitem[\protect\citeauthoryear{Zhou, Shi, Zhang  \& Wang}{Zhou et~al.}{2021}]{zhou_extremely_2021}
Zhou L.,  Shi Y.,  Zhang Z.-Y.,   Wang J.,  2021, \mn@doi [Astronomy \& Astrophysics] {10.1051/0004-6361/202039033}, 653, L10

\makeatother
\end{thebibliography}
\label{lastpage}
\clearpage
\appendix

\section{Resolution tests} \label{ResTests}
As mentioned in section \ref{Simulations}, we ran resolution tests for our galaxy models. We ran low-resolution versions of each galaxy model, with 8x lower resolution, and a high-resolution version of m3e10 with 8x higher resolution.

We have shown results for how the CO emission traces the H$_2$ gas with the CO-dark ($W_{\hbox{\textsc{co}}}$> 0.1 K km s$^{-1}$) H$_2$ gas percentages in table \ref{COdarkTable} for our standard resolution simulations with non-equilibrium chemistry and post-processed equilibrium chemistry. We can now look at how these percentages change when we alter the resolution of the simulation in Table \ref{COdarkResTable}, where we show the percentage of CO-dark ($W_{\hbox{\textsc{co}}}$> 0.1 K km s$^{-1}$) H$_2$ gas within the low-resolution simulations, and the high-resolution version of m3e10. Unlike for our non-equilibrium and equilibrium simulations, we do not run resolution tests for the m3e11 gas fraction variants.

\begin{table} 
\begin{minipage}{84mm}
\centering
\caption{The percentage of H$_2$ gas which is considered CO-dark ($W_{\mbox{\textsc{co}}}$ > 0.1 K km s$^{-1}$) within each galaxy at 500 Myr for our fiducial model with non-equilibrium chemistry (Fiducial), the low resolution (low\_res) and high resolution simulations (hi\_res).} \label{COdarkResTable}
\begin{tabular}{cccc}
  \hline
  Name & low\_res & Fiducial & hi\_res \\
  \hline
  m1e10 & 100\% & 97.6\% &-\\
  m3e10 & 77.3\% & 94.4\% & 67.8\% \\
  m1e11 & 54.3\% & 71.7\% & -\\
  m3e11 & 47.9\% & 54.0\% & -\\
  m1e12 & 37.2\% & 40.4\% & -\\
\hline
\end{tabular}
\end{minipage}
\end{table} 

Using these results, we can see how resolution can affect the amount of CO-dark gas within these simulations, as within the m1e10 galaxy we get no CO emission above our 0.1 K km s$^{-1}$ detection threshold, and so the entire galaxy is classed as CO-dark gas. Within the low-resolution versions of m3e10 and m1e11 we see that we are predicting \textasciitilde20\% less CO-dark gas compared to our non-equilibrium simulations. We also see that there is no general trend of CO-dark gas with resolution for our m3e10 galaxies, which suggests that the differences in these dwarf galaxies are driven partly by stochastic effects. Finally, in the low-resolution versions of m3e11 and m1e12, we see that we nearly match the amount of CO-dark gas within our non-equilibrium simulations in our low-resolution versions, differing by only 2\% to 6\%.

We can look at how resolution can change our H$_2$ mass, as well as the luminosity of our simulated CO emission, and the resulting $X_{\hbox{\textsc{co}}}$ calculated from the two. We compare these in Table \ref{ResRatioTable}, where we can show the effects of resolution on our simulations for our low-resolution simulations, as well as our high resolution run of m3e10. We average over all 5 snapshots for each galaxy to reduce the effect the bursty nature of the dwarf galaxies has on the results.

\definecolor{pos1_1}{rgb}{0.24229,0.45445,0.57568}
\definecolor{pos1_2}{rgb}{0.85418,0.89501,0.91834}
\definecolor{pos1_3}{rgb}{0.81838,0.76648,0.67567}
\definecolor{pos1_4}{rgb}{0.85475,0.81325,0.74063}
\definecolor{pos1_5}{rgb}{0.93242,0.95134,0.96215}
\definecolor{pos1_6}{rgb}{0.9409,0.95745,0.9669}


\definecolor{pos2_1}{rgb}{0.24229,0.45445,0.57568}
\definecolor{pos2_2}{rgb}{0.72973,0.80541,0.84865}
\definecolor{pos2_3}{rgb}{0.9454,0.96069,0.96942}
\definecolor{pos2_4}{rgb}{0.96988,0.97832,0.98314}
\definecolor{pos2_5}{rgb}{0.95028,0.9642,0.97216}
\definecolor{pos2_6}{rgb}{0.96773,0.97677,0.98193}


\definecolor{pos3_1}{rgb}{0.46206,0.30836,0.03939}
\definecolor{pos3_2}{rgb}{0.9214,0.9434,0.95598}
\definecolor{pos3_3}{rgb}{0.55024,0.67617,0.74814}
\definecolor{pos3_4}{rgb}{0.67927,0.76907,0.82039}
\definecolor{pos3_5}{rgb}{0.99166,0.98927,0.9851}
\definecolor{pos3_6}{rgb}{0.98669,0.98289,0.97623}

\begin{table}
\begin{minipage}{84mm}
\centering
\caption{Ratios of CO emission line luminosity, H$_2$ mass, and $X_{\mbox{\textsc{co}}}$, $Y_{\rm{res}} / Y_{\rm{noneq}}$ for the resolution tests compared to the non-equilibrium fiducial model, comparing the average across both simulations. ResType tells us whether we are comparing with the low resolution (low\_res) or high resolution (hi\_res) version of each model. Values highlighted in red and blue correspond to an enhancement and reduction, respectively.} \label{ResRatioTable}
\begin{tabular}{ccccc}
  \hline
   & & \multicolumn{3}{c}{$Y_{\rm{res}} / Y_{\rm{noneq}}$} \\ 
  \cline{3-5} 
  Galaxy & ResType & [$L_{\mbox{\textsc{co}}}$] & [$M_{\mbox{\textsc{h}}_{\oldstylenums{2}}}$] & [$X_{\mbox{\textsc{co}}}$] \\ 
\hline
m1e10 & low\_res& \cellcolor{pos1_1}0.00756 & \cellcolor{pos2_1}0.16099 & \cellcolor{pos3_1}21.30591 \\
m3e10 & low\_res& \cellcolor{pos1_2}0.6316& \cellcolor{pos2_2}0.48052 & \cellcolor{pos3_2}0.76079 \\
m3e10 & hi\_res& \cellcolor{pos1_3}2.29731 & \cellcolor{pos2_3}0.82075 & \cellcolor{pos3_3}0.35727 \\
m1e11 & low\_res & \cellcolor{pos1_4}2.03749 & \cellcolor{pos2_4}0.89249 & \cellcolor{pos3_4}0.43803 \\
m3e11 & low\_res& \cellcolor{pos1_5}0.78719 & \cellcolor{pos2_5}0.83411 & \cellcolor{pos3_5}1.0596\\
m1e12 & low\_res& \cellcolor{pos1_6}0.80879 & \cellcolor{pos2_6}0.88569 & \cellcolor{pos3_6}1.09508\\
\hline
\end{tabular}
\end{minipage}
\end{table} 

In our dwarf galaxy m1e10, we see a reduction on all values when compared to our fiducial model values. However, as we mentioned prior, m1e10 exhibits extremely bursty star formation histories and gas fractions, which we see when we compare our fiducial model to our equilibrium simulations as well. We see in Table \ref{COdarkResTable} that we see no CO emission in our snapshot taken at time $t = 500$ Myr for m1e10, meaning we are producing little-to-no CO emission at times, causing the exaggerated reductions. In our high-mass galaxies, m3e11 and m1e12, we see that there is only a small fluctuation between our fiducial model and the low-resolution tests for $X_{\hbox{\textsc{co}}}$, and a reduction of $\approx$ 12\% to 22\% for the $L_{\hbox{\textsc{co}}}$ and $M_{\mbox{\textsc{h}}_{\oldstylenums{2}}}$. For our high-resolution simulation of m3e10, we see that $L_{\hbox{\textsc{co}}}$ is enhanced by over a factor of 2, and $M_{\mbox{\textsc{h}}_{\oldstylenums{2}}}$ is reduced by \textasciitilde22\%, causing a greater reduction in the final $X_{\hbox{\textsc{co}}}$ down to 35\% the value we see in our fiducial simulations. Like m1e10, m3e10 also exhibits bursty star formation, which may contribute to this result.

\begin{figure}%
    \includegraphics*[width=\columnwidth]{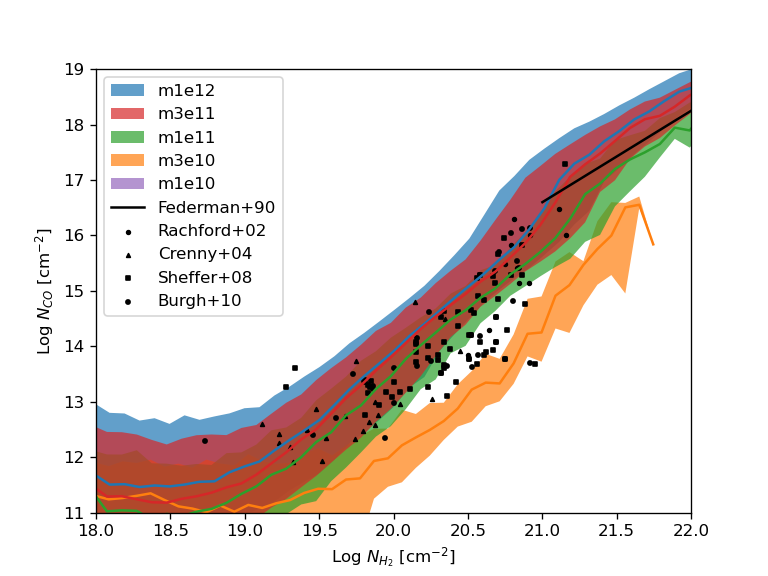}
    \caption{Relation between CO and H$_2$ column densities for the range of halo masses of our low-resolution tests. The shaded coloured regions indicate the 15th to 85th percentile within each H$_2$ column density bin, while the solid curves show the median, for each halo mass at time $t =$ 500 Myr. The black points included show observational data from UV absorption lines in the Milky Way from  \citet{federman_modeling_1990,rachford_far_2002,crenny_reanalysis_2004,sheffer_ultraviolet_2008}, and \citet{burgh_atomic_2010}.} \label{lowrescols}
\end{figure}

We can also plot the relation between $N_{\mbox{\textsc{co}}}$ vs $N_{\mbox{\textsc{h}}_{\oldstylenums{2}}}$ for our resolution tests in Figure \ref{lowrescols}, showing how our low-resolution tests compare to the observations as we did for our fiducial models in Section \ref{ObsCols}. We see that our low-resolution models also match the observed relation within the Milky Way between column densities. As our low-resolution version of our Milky Way-mass galaxy m1e12 at time $t = 500$ Myr is closer to solar metallicity than our non-equilibrium fiducial model, we do see that it is \textasciitilde0.25dex below our fiducial model. 

\begin{figure}%
  \includegraphics*[width=\columnwidth]{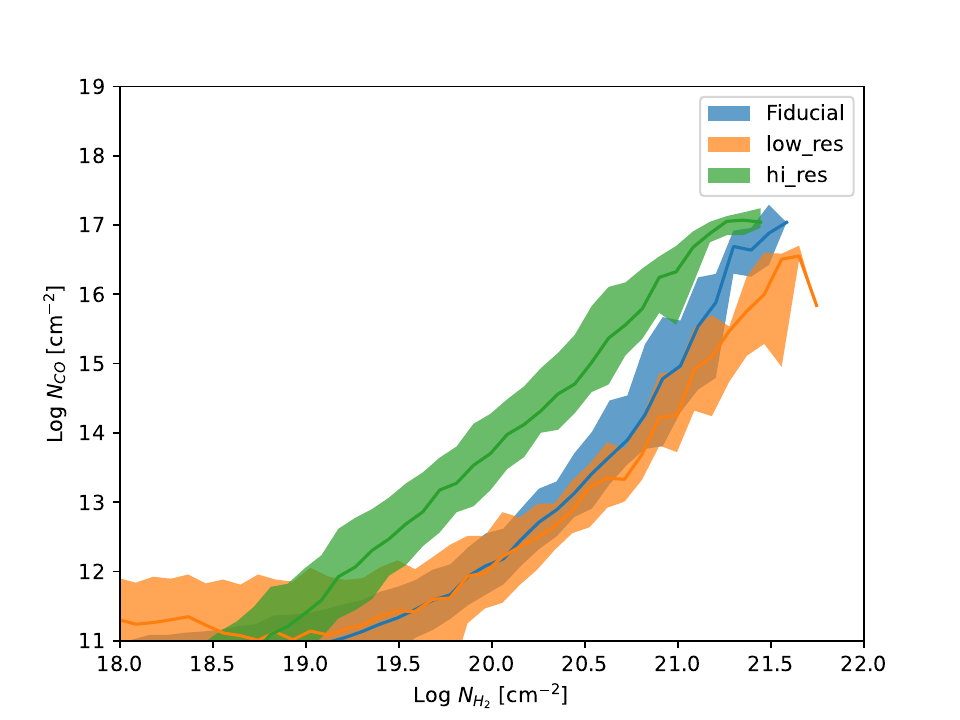}
    \caption{Relation between CO and H$_2$ column densities for the range of our m3e10 simulations, with our fiducial model, and our low- and high-resolution test. The shaded coloured regions indicate the 15th to 85th percentile within each H$_2$ column density bin, while the solid curves show the median, for each simulation at time $t =$ 500Myr.} \label{m3e10Cols}
\end{figure}

We also run a version of this plot comparing both our resolution tests for m3e10 to our fiducial non-equilibrium model in Figure \ref{m3e10Cols}. We see that for a fixed H$_2$ column density at $N_{\mbox{\textsc{h}}_{\oldstylenums{2}}}$\textasciitilde10$^{20}$ cm$^{-2}$, we produce CO column density an order of magnitude above our fiducial model in our high-resolution version of m3e10. However, at higher H$_2$ column densities ($N_{\mbox{\textsc{h}}_{\oldstylenums{2}}}$>\textasciitilde10$^{21}$ cm$^{-2}$), we see that our fiducial model agrees more with our high-resolution version over the low-resolution. We also see in Table \ref{ResRatioTable} that whilst we are producing a higher ratio of CO to H$_2$ in this resolution test, that our $L_{\mbox{\textsc{co}}}$ is only affected by a factor of 2.3, and the final $X_{\mbox{\textsc{co}}}$ is affected by a factor of 3, which both are similar to other resolution tests, such as our m1e11 low-resolution test. This smaller difference in the overall CO luminosity is because it is dominated  by the high column density regions.

\begin{figure} 
    \includegraphics*[width=\columnwidth]{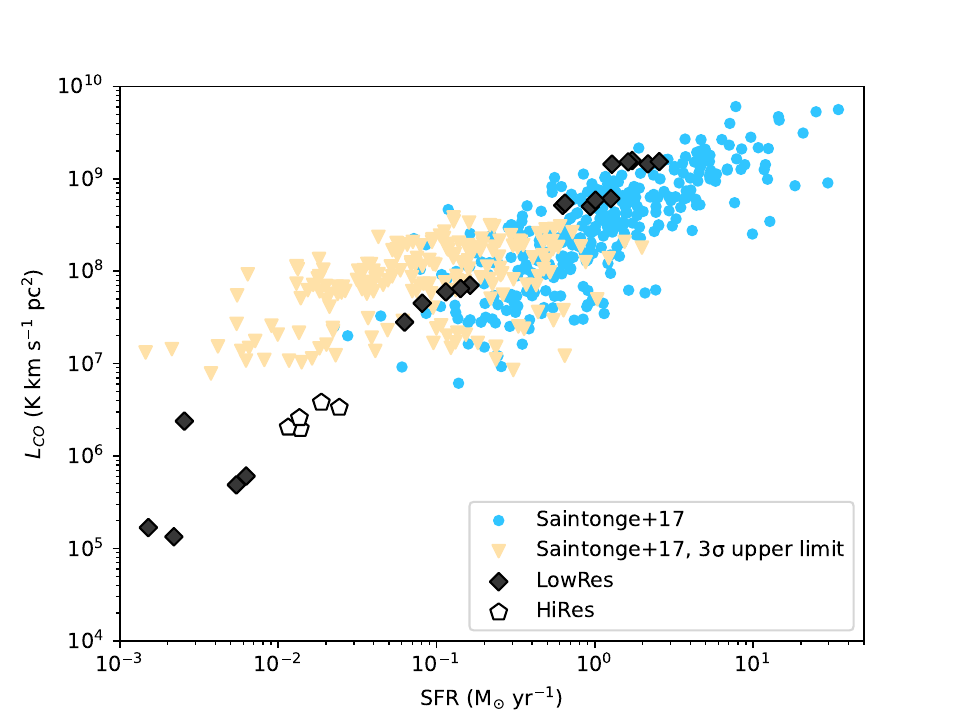}
    \caption{$L_{\mbox{\textsc{co}}}$ against SFR for each galaxy for each snapshot, with the low-resolution tests in grey diamonds, and the m3e10 high-resolution tests in white pentagons. The xCOLD GASS survey points are included in the blue, with the measurements of galaxies with no CO (1-0) emission detected shown in the yellow arrows, which are set as a 3$\sigma$ upper limit on the CO luminosity.} \label{ResLCO}
\end{figure}

We can also compare our resolution tests to the observed CO luminosities of the xCOLD GASS survey in Figure \ref{ResLCO}. We see that our mid-to-high mass runs (with high SFR) match observed luminosities of objects with similar SFR in the xCOLD GASS survey. Our low-mass galaxies, and therefore our high-resolution runs of m3e10, lie below the xCOLD GASS survey, as we expect and discuss in Section \ref{xCOLDSec}.

\begin{figure} 
    \includegraphics*[width=\columnwidth]{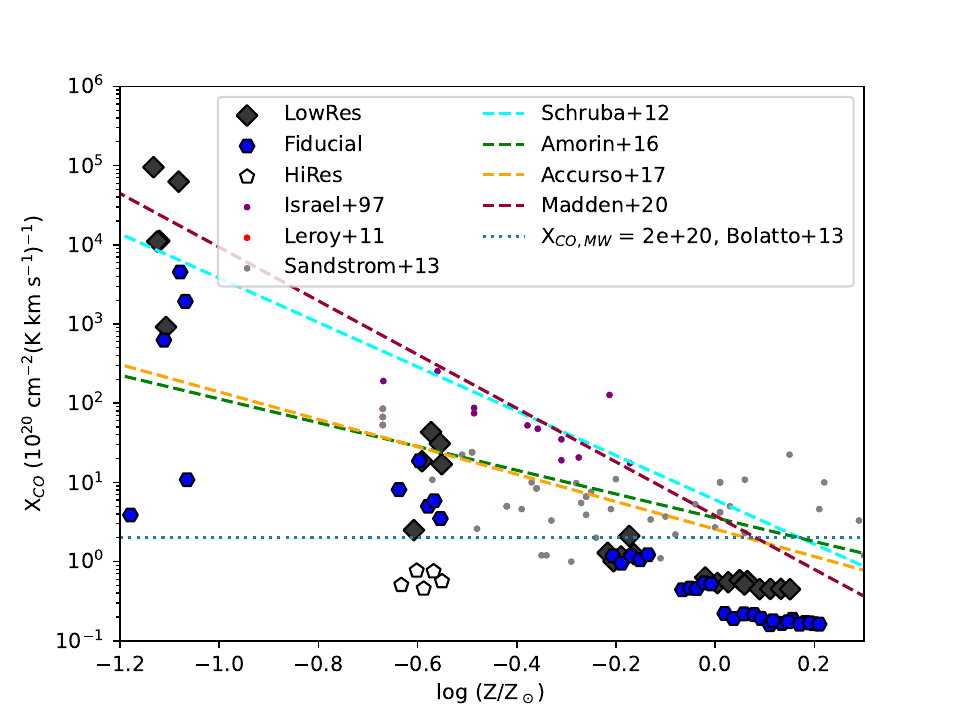}
    \caption{$X_{\mbox{\textsc{co}}}$ against gas-phase metallicity for each snapshot with the low-resolution tests in grey, and the m3e10 high-resolution tests in white. We also plot our non-equilibrium fiducial results in blue hexagons. We plot dust-based observational measurements from \citet{israel_h2_1997,leroy_co--h_2011} and \citet{sandstrom_co--h_2013}, as well as the $X_{\mbox{\textsc{co}}}$ for the Milky-Way with the blue-dotted line. We also plot metallicity-dependent relations for the conversion factor with coloured dashed lines from \citet{schruba_low_2012,amorin_molecular_2016,accurso_deriving_2017}, and \citet{madden_tracing_2020}. We show metallicity when compared to solar metallicity, Z$_\odot$.} \label{ResXCO}
\end{figure}

We also compare our resolution tests against the relation between $X_{\hbox{\textsc{co}}}$ and metallicity in Figure \ref{ResXCO}, as well as to our fiducial model. We see that our resolution tests match the observed relation between $X_{\hbox{\textsc{co}}}$ and metallicity. 

As we see in our results, we see multiple differences in the high-resolution version of m3e10 when compared to our low-resolution and fiducial models. However, we believe these differences are driven by the gas content and star formation history of this high-resolution model, which we can see in Figure \ref{ResLCO}, where our high-resolution m3e10 has a much higher SFR than the low-resolution version. We also note that our non-equilibrium fiducial model and high-resolution test differ by a factor of 2 for the CO luminosity, and the conversion factor. This is to be expected, as we see large fluctuations from snapshot to snapshot for these dwarf galaxies in our simulations, which can be seen in both Figure \ref{ResLCO} and Figure \ref{ResXCO} where we see large fluctuations within snapshots of the same resolution. Therefore, we also believe part of these differences to be due to the bursty nature of the dwarf galaxies. Further study into the dwarf regime is required to understand this further.

\bsp	
\end{document}